\documentclass[runningheads,a4paper]{llncs}
\usepackage{amssymb}
\usepackage{graphicx}

\usepackage{amsmath}
\usepackage{listings}

\usepackage{url}
\urldef{\mailsa}\path|{shaojianxiong, qin_yu, feng}@tca.iscas.ac.cn|
\urldef{\mailsb}\path|feng@tca.iscas.ac.cn|
\urldef{\mailsc}\path|qin_yu@tca.iscas.ac.cn|

\newcommand{\keywords}[1]{\par\addvspace\baselineskip
\noindent\keywordname\enspace\ignorespaces#1}

\begin{document}

\mainmatter

\title{Computational Soundness Results for\\ Stateful Applied $\pi$ Calculus}
\titlerunning{Computational Soundness Results for Stateful Applied $\pi$ Calculus}

\author{Jianxiong Shao%
\thanks{The research presented in this paper is supported by the National
Basic Research Program of China (No. 2013CB338003) and National
Natural Science Foundation of China (No. 91118006, No.61202414).}
\and Yu Qin\and Dengguo Feng}

\authorrunning{Jianxiong Shao\and Dengguo Feng\and Yu Qin}

\institute{Trusted Computing and Information Assurance Laboratory,\\
Institute of Software, Chinese Academy of Sciences, Beijing, China\\
\mailsa\\
%\url{http://www.springer.com/lncs}
}

\toctitle{Computational Soundness Results for Stateful Applied $\pi$ Calculus}
\tocauthor{Jianxiong Shao, Yu Qin, Dengguo Feng}

\maketitle

\begin{abstract}
%\boldmath

In recent years, many researches have been done to
establish symbolic models of stateful protocols. Two works among
them, the SAPIC tool and StatVerif tool, provide a high-level
specification language and an automated analysis. Their language, the
stateful applied $\pi$ calculus, is extended from the applied $\pi$
calculus by defining explicit state constructs. Symbolic
abstractions of cryptography used in it make the analysis amenable
to automation. However, this might overlook the attacks based on the
algebraic properties of the cryptographic algorithms. In our paper,
we establish the computational soundness results for stateful
applied $\pi$ calculus used in SAPIC tool and StatVerif tool.

In our approach, we build our results on the CoSP framework. For
SAPIC, we embed the non-monotonic protocol states into the CoSP
protocols, and prove that the resulting CoSP protocols are
efficient. Through the embedding, we provide the computational
soundness result for SAPIC (by Theorem 1). For StatVerif, we encode
the StatVerif process into a subset of SAPIC process, and obtain the
computational soundness result for StatVerif (by Theorem 2). Our
encoding shows the differences between the semantics of the two
languages. Our work inherits the modularity of CoSP, which allows
for easily extending the proofs to specific cryptographic
primitives. Thus we establish a computationally sound automated
verification result for the input languages of SAPIC and StatVerif
that use public-key encryption and signatures (by Theorem 3).

\end{abstract}

\section{Introduction}

Manual proofs of security protocols that rely on cryptographic
functions are complex and known to be error-prone. The complexity
that arises from their distributed nature motivates the researches
on automation of proofs. In recent years, many efficient
verification tools (\cite{Proverif,AVISPA,Maude}) have been
developed to prove logical properties of protocol behaviors. To
eliminate the inherent complexity of the cryptographic operations in
formal analysis, these verification tools abstract the cryptographic
functions as idealized symbolic terms that obey simple cancelation
rules, i.e., the so-called Dolev-Yao models
(\cite{DYModel,DYModel2}). Unfortunately, these idealizations also
abstract away from the algebraic properties a cryptographic
algorithm may exhibit. Therefore a symbolic formal analysis may omit
attacks based on these properties. In other words, symbolic security
does not immediately imply computational security. In order to
remove this limitation, the concept of Computational
Soundness (CS) is introduced in \cite{Reconcil}. From
the start, a large number of CS results over the past decade were
made to show that many of the Dolev-Yao models are sound with
respect to actual cryptographic realizations and security
definitions (see, e.g.,
\cite{Completing,Soundness,Computationally,Computationally06,CoSP,Computational,Composable,Deduction,Computational12}).

More recently, formal analysis methods have been applied to stateful
protocols, i.e., protocols which require \emph{non-monotonic global
state} that can affect and be changed by protocol runs. Stateful
protocols can be used to model hardware devices that have some
internal memory and security APIs, such as the RSA PKCS\#11, IBM's
CCA, or the trusted platform module. There are many formal methods
that have been used to establish symbolic model of stateful
protocols
(\cite{stateful14,StateStrand,tamarin,stateR,SAPIC,Statverif,Abstraction}).
Two works among them, the SAPIC
tool \cite{SAPIC} and StatVerif tool \cite{Statverif}, can provide an automated analysis of stateful
protocols. Their language, the stateful applied $\pi$ calculus, is
extended from the applied $\pi$ calculus \cite{APicalculus} by
defining constructs for explicitly manipulating global state. One
advantage of the stateful applied $\pi$ calculus is that it provides
a high-level specification language to model stateful protocols. Its
syntax and semantics inherited from the applied $\pi$ calculus can
arguably ease protocol modeling. Another advantage is that the
formal verification can be performed automatically by these tools.

However, no CS works have been done for the stateful applied $\pi$
calculus. Although there are many for the original applied $\pi$
calculus, e.g., see \cite{CoSP,Computational12,CS14}. Our purpose is
to establish the CS results for the input languages of the two
verification tools SAPIC and StatVerif. With our results, we can
transform their symbolically automated verification results of
stateful protocols (with some restrictions) to the computationally sound
one with respect to actual cryptographic realizations and security
definitions. We want to establish the CS results directly for
the input languages of SAPIC and StatVerif. To achieve this, we
choose to embed them into the CoSP work \cite{CoSP}, a general
framework for conceptually modular CS proofs. Since the stateful
applied $\pi$ calculus used in SAPIC and StatVerif are slightly
different, in the following we call the former SAPIC calculus and
the latter StatVerif calculus.

\noindent \textbf{Our work}. We present two CS results respectively
for the stateful applied $\pi$ calculus used in SAPIC tool and
StatVerif tool. In our approach, we first provide the method to
embed SAPIC calculus into the CoSP framework. Note that the CoSP
framework does not provide explicit state manipulation. We need to
embed the complex state constructs of stateful applied $\pi$
calculus into the CoSP protocols and make sure that the resulting
CoSP protocol is efficient. By the embedding, we prove that the CS
result of applied $\pi$ calculus implies that of SAPIC calculus (by
Theorem 1). For StatVerif, we provide an encoding of StatVerif
processes into a subset of SAPIC processes and build the CS result
of StatVerif calculus (by Theorem 2). Our encoding shows the
differences between the semantics of these two languages. Finally,
we establish a computationally sound automated verification result
for the input languages of SAPIC and StatVerif that use public-key
encryption and signatures (by Theorem 3).

For SAPIC, we use the calculus proposed by \cite{SAPIC} as the SAPIC
calculus. It extends the applied $\pi$ calculus with two kinds of
state: the functional state and the multiset state. We set two
restrictions respectively for the pattern matching in the input
constructs and for the multiset state constructs. They are necessary
for the computational execution model. We embed the SAPIC calculus
into the CoSP framework. The two kinds of state are encoded into the
CoSP protocol state (as part of the CoSP node identifiers). We have
met two challenges in the embedding. First is for the functional
state. If we encode them directly as $\pi$-terms, the resulting CoSP
protocol is not efficient. Thus we transform them into the CoSP
terms which are treated as black boxes by CoSP protocols. The second
problem is for the encoding of multiset state. By our restriction of
multiset state constructs, we can transform the arguments of facts
into CoSP terms and limit the growth of the size of multiset state.
We also provide an efficient CoSP sub-protocol to implement the
pattern matching in the multiset state constructs. At last, we prove
that our embedding is an efficient and safe approximation of the
SAPIC calculus, and build the CS result of SAPIC calculus upon that
of applied $\pi$ calculus (by Theorem 1).

For StatVerif, we use the calculus proposed by \cite{Statverif} as
the StatVerif calculus. It has minor differences to SAPIC calculus.
We first provide an encoding of the StatVerif processes into a
subset of SAPIC processes. Then we prove that by using SAPIC trace
properties our encoding is able to capture secrecy of stateful
protocols. With the CS result of SAPIC, we can directly obtain the
CS result of StatVerif calculus (by Theorem 2). Our encoding
shows the differences between the semantics of state constructs in
these two calculi.

Note that our contribution is a soundness result for the execution
models that can manipulate state, rather than a soundness result for
any new cryptographic primitives. The advantage of our CS result is
its extensibility, since we build it on the CoSP framework and
involve no new cryptographic arguments. It is easy to extend our
proofs to additional cryptographic abstractions phrased in CoSP
framework. Any computationally sound implementations for applied
$\pi$ calculus that have been proved in CoSP framework can be
applied to our work. To explain its extendibility, we establish a
computationally sound automated verification result for the input
languages of SAPIC and StatVerif that use public-key encryption and
signatures (by Theorem 3). We have verified the classic left-or-right
protocol presented in \cite{Statverif} by using these tools in a
computationally sound way to show the usefulness of our result.

The paper is organized as follows. In Section 2 we give a brief
introduction to the CoSP framework and the embedding of applied
$\pi$ calculus. In Section 3 and Section 4 we respectively show the
CS results of stateful applied $\pi$ calculus in SAPIC and StatVerif
work. Section 5 contains a case study of the CS result of public-key
encryption and signatures. We conclude in Section 6.

\section{Preliminaries}

\subsection{CoSP Framework}

Our CS results are formulated within CoSP \cite{CoSP}, a framework
for conceptually modular CS proofs. It decouples the treatment of
cryptographic primitives from the treatment of calculi. The results
in \cite{Computational12} and \cite{CS14} have shown that CoSP
framework is capable of handling CS with respect to trace properties
and uniformity for ProVerif. Several calculi such as the applied
$\pi$ calculus and RCF can be embedded into CoSP (\cite{CoSP,CS10})
and combined with CS results for cryptographic primitives. In this
subsection, we will give a brief introduction to the CoSP framework.

CoSP provides a general symbolic model for abstracting cryptographic
primitives. It contains some central concepts such as constructors,
destructors, and deduction relations.

\noindent\textbf{Definition 1 (Symbolic Model)}. A symbolic model
$\textbf{M}=(\textbf{C},\textbf{N},\textbf{T},\textbf{D},\vdash)$
consists of a set of constructors \textbf{C}, a set of nonces
\textbf{N}, a message type \textbf{T} over \textbf{C} and \textbf{N}
with $\textbf{N}\subseteq \textbf{T}$, a set of destructors
\textbf{D} over \textbf{T}, and a deduction relation $\vdash$ over
\textbf{T}. A constructor $C/n \in \textbf{C}$ is a symbol with
(possible zero) arity. A nonce $N \in \textbf{N}$ is a symbol with
zero arity. A message type \textbf{T} is a set of terms over
constructors and nonces. A destructor $D/n \in \textbf{D}$ of arity
$n$ over a message type \textbf{T} is a partial map $\textbf{T}^n
\rightarrow \textbf{T}$. If $D$ is undefined on a list of message
$\underline{t}=(t_1,\cdots,t_n)$, then $D(\underline{t})=\bot$.

To unify notation of constructor or destructor $F/n \in
\textbf{C}\cup \textbf{D}$ and nonce $F \in \textbf{N}$, we define
the partial function $eval_F: \textbf{T}^n \rightarrow \textbf{T}$,
where $n=0$ for the nonce, as follows: If $F$ is a constructor,
$eval_F(\underline{t}):=F(\underline{t})$ if $F(\underline{t})\in
\textbf{T}$ and $eval_F(\underline{t}):=\bot$ otherwise. If $F$ is a
nonce, $eval_F():=F$. If $F$ is a destructor,
$eval_F(\underline{t}):=F(\underline{t})$ if $F(\underline{t})\neq
\bot$ and $eval_F(\underline{t}):= \bot$ otherwise.

A \emph{computational implementation} $A$ of a symbolic model
$\textbf{M}$ is a family of algorithms that provide computational
interpretations to constructors, destructors, and specify the
distribution of nonces.

A \emph{CoSP protocol} $\Pi$ is a tree with labelled nodes and
edges. Each node has a unique identifier. It distinguishes 4 types
of nodes. \emph{Computation nodes} describe constructor
applications, destructor applications, and creations of nonce.
\emph{Output} and \emph{input nodes} describe communications with
the adversary. \emph{Control nodes} allow the adversary to choose
the control flow of the protocol. The computation nodes and input
nodes can be referred to by later computation nodes or output nodes.
The messages computed or received at these earlier nodes are then
taken as arguments by the later constructor/destructor applications
or sent to the adversary. A CoSP protocol is \emph{efficient} if it
satisfies two conditions: for any node, the length of the identifier
is bounded by a polynomial in the length of the path (including the
total length of the edge-labels) from the root to it; there is a
deterministic polynomial-time algorithm that, given the labels of
all nodes and edges on the path to a node, computes the node's
identifier.

Given an efficient CoSP protocol $\Pi$, both its \emph{symbolic} and
\emph{computational executions} are defined as a valid path through
the protocol tree. In the symbolic execution, the computation nodes
operate on terms, and the input (resp. output) nodes receive (resp.
send) terms to the symbolic adversary. The successors of control
nodes are chosen by the adversary. In the computational execution,
the computation nodes operate on bitstrings by using a computational
implementation $A$, and the input (resp. output) nodes receive
(resp. send) bitstrings to the polynomial-time adversary. The
successors of control nodes are also chosen by the adversary. The
symbolic (resp. computational) \emph{node trace} is a list of node
identifiers if there is a symbolic (resp. computational) execution
path with these node identifiers.

\noindent\textbf{Definition 2 (Trace Property)}. A trace property
$\wp $ is an efficiently decidable and prefix-closed set of (finite)
lists of node identifiers. Let
$\textbf{M}=(\textbf{C},\textbf{N},\textbf{T},\textbf{D},\vdash)$ be
a symbolic model and $\Pi$ be an efficient CoSP protocol. Then $\Pi$
symbolically satisfies a trace property $\wp $ in \textbf{M}
\emph{iff} every symbolic node trace of $\Pi$ is contained in $\wp
$. Let $A$ be a computational implementation of \textbf{M}. Then
$(\Pi, A)$ computationally satisfies a trace property $\wp $ in
\textbf{M} \emph{iff} for all probabilistic polynomial-time
interactive machines $\mathcal {A}$, the computational node trace is
in $\wp$ with overwhelming probability.

\noindent\textbf{Definition 3 (Computational Soundness)}. A
computational implementation $A$ of a symbolic model
$\textbf{M}=(\textbf{C},\textbf{N},\textbf{T},\textbf{D},\vdash)$ is
computationally sound for a class $P$ of CoSP protocols \emph{iff}
for every trace property $\wp $ and for every efficient CoSP
protocol $\Pi \in P$, we have that $(\Pi,A)$ computationally
satisfies $\wp$ whenever $\Pi$ symbolically satisfies $\wp$.

\subsection{Embedding Applied $\pi$ Calculus into CoSP Framework}

Stateful applied $\pi$ calculus is a variant of applied $\pi$
calculus. We need to review the original applied $\pi$ calculus
first. We provide its syntax in \textbf{Table \ref{standard}}. It
corresponds to the one considered in \cite{CoSP}.

\begin{table}
\begin{center}
  \caption{Syntax of applied $\pi$ calculus}  \label{standard}%
    \begin{tabular}{llll}
\hline
    $\langle M,N\rangle ::=$ & terms&                                                     $\langle P,Q\rangle ::=$ & processes\\
\quad  \quad   $a, b, m, n, ...$           & names&                             \quad  \quad    $0$                            & nil \\
\quad  \quad   $x, y, z, ...$                    & variables &                         \quad  \quad   $P|Q $                          & parallel  \\
\quad \quad    $f(M_1, ..., M_n)$                & constructor applications  &        \quad \quad    $!P $                           & replication\\
&       &                                                                              \quad  \quad   $ \nu n;P $                    & restriction\\
$D::=$ &  destructor terms  &                                                          \quad  \quad   out$(M,N);P$                   & output\\
\quad \quad $M, N, ...$                          & terms &                             \quad  \quad in$(M,x);P $                     & input\\
\quad \quad $d(D_1, ..., D_n)$                   & destructor applications &            \quad \quad let $x=D$ in $P$ else $Q$      & let\\
\quad \quad $f(D_1, ..., D_n)$                   & constructor applications  &        \quad  \quad     event $e;P $                   & event\\
  \hline
    \end{tabular}%
\end{center}
\end{table}%

In the following, we call the terms in process calculus the
$\pi$-terms and terms in CoSP the CoSP-terms, in order to avoid
ambiguities. It is similar for the other homonyms such as
$\pi$-constructors. We will use $fn(P)$ (resp. $fv(P)$) for free
names (resp. free variables) in process $P$, i.e., the names (resp.
variables) that are not protected by a name restriction (resp. a let
or an input). The notations can also be applied to terms in process.
We call a process closed or a term ground if it has no free
variables.

The calculus is parameterized over a set of $\pi$-constructors
$\textbf{C}_\pi$, a set of $\pi$-destructors $\textbf{D}_\pi$, and
an equational theory $E$ over ground $\pi$-terms. It requires that
the equational theory is compatible with the $\pi$-constructors and
$\pi$-destructors as defined in \cite{CoSP}. The symbolic model of
applied $\pi$-calculus can be embedded into the CoSP framework.

\noindent\textbf{Definition 4 (Symbolic Model of the Applied $\pi$
Calculus)}. For a $\pi$-destructor $d\in \textbf{D}_\pi$, the
CoSP-destructor $d'$ is defined by
$d'(\underline{t}):=d(\underline{t}\rho)\rho^{-1}$ where $\rho$ is
any injective map from the nonces occurring in the CoSP-terms
$\underline{t}$ to names. Let $\textbf{N}_E$ for adversary nonces
and $\textbf{N}_P$ for protocol nonces be two countably infinite
sets. The symbolic model of the applied $\pi$ calculus is given by
$\textbf{M}=(\textbf{C},\textbf{N},\textbf{T},\textbf{D},\vdash)$,
where $\textbf{N}:=\textbf{N}_E\cup\textbf{N}_P$,
$\textbf{C}:=\textbf{C}_\pi$,
$\textbf{D}:=\{d'|d\in\textbf{D}_\pi\}$, and where $\textbf{T}$
consists of all terms over \textbf{C} and \textbf{N}, and where
$\vdash$ is the smallest relation such that $m\in S \Rightarrow
S\vdash m$, $N\in \textbf{N}_E\Rightarrow S\vdash N$, and such that
for any $F\in \textbf{C}\cup\textbf{D}$ and any
$\underline{t}=(t_1,...,t_n)\in\textbf{T}^n$ with
$S\vdash\underline{t}$ and $eval_F(\underline{t})\neq\bot$, we have
$S\vdash eval_F(\underline{t})$.

The if-statement can be expressed using an additional destructor
$equal$, where $equal(M,N)=M$ if $M=_E N$ and $equal(M,N)=\bot$
otherwise. We always assume $equal\in \textbf{D}_\pi$. The
destructor $equal'$ induces an equivalence relation $\cong$ on the
set of CoSP-terms with $x\cong y$ \emph{iff} $equal'(x,y)\neq\bot$.

For the symbolic model, we can specify its computational
implementation $A$. It assigns the deterministic polynomial-time
algorithms $A_f$ and $A_d$ to each $\pi$-constructors and
$\pi$-destructors, and chooses the nonces uniformly at random.

We introduce some notations for the definitions of computational and
symbolic $\pi$-executions. Given a ground destructor CoSP-term $D'$,
we can evaluate it to a ground CoSP-term $\text{eval}^{CoSP}(D')$ by
evaluating all CoSP-destructors in the arguments of $D'$. We set
$\text{eval}^{CoSP}(D'):=\bot$ \emph{iff} any one of the
CoSP-destructors returns $\bot$. Given a destructor $\pi$-term $D$,
an assignment $\mu$ from $\pi$-names to bitstrings, and an
assignment $\eta$ from variables to bitstrings with $fn(D) \subseteq
dom(\mu)$ and $fv(D) \subseteq dom(\eta)$, we can computationally
evaluate $D$ to a bitstring ceval$_{\eta,\mu}D$. We set
$\text{ceval}_{\eta,\mu}D:=\bot$ if the application of one of the
algorithms $A^\pi_f$ or $A^\pi_d$ fails. For a partial function $g$,
we define the function $f:=g \cup \{a := b\}$ with
$dom(f)=dom(g)\cup\{a\}$ as $f(a):=b$ and $f(x):=g(x)$ for $x\neq
a$.

%In the following, we say a nondeterministic interactive machine
%$\mathcal{A}$ is a Dolev-Yao adversary if the following holds in an
%interaction with any interactive machine $\mathcal{M}$ in each step
%of the interaction: Let $K$ be the set of all CoSP-terms sent by
%$\mathcal{M}$ up to the current step. Let $m$ be the term sent by
%$\mathcal{A}$ in the current step. Then $K\vdash m$.

The computational and symbolic execution models of a $\pi$-process
are defined in \cite{CoSP} by using evaluation contexts where the
holes only occur below parallel compositions. The adversary is
allowed to determine which process in parallel should be proceeded
by setting the evaluation context for each step of proceeding. The
execution models of $\pi$ calculus are defined as follows. We take
the writing way in \cite{CoSP} and mark the symbolic execution model
by $[\![...]\!]$.

\noindent\textbf{Definition 5 $[\![$6$]\!]$ (Computational
$[\![$Symbolic$]\!]$ Execution of $\pi$ Calculus)}.  Let $P_0$ be a
closed process (where all bound variables and names are renamed such
that they are pairwise distinct and distinct from all unbound ones).
Let $\mathcal {A}$ be an interactive machine called the adversary.
$[\![$For the symbolic model, $\mathcal {A}$ only sends message $m$
if $K \vdash m$ where $K$ are the messages sent to $\mathcal {A}$ so
far.$]\!]$ We define the computational $[\![$symbolic$]\!]$
execution of $\pi$ calculus as an interactive machine
$Exec_{P_0}(1^k)$ that takes a security parameter $k$ as argument
$[\![$interactive machine $SExec_{P_0}$ that takes no argument$]\!]$
and interacts with $\mathcal {A}$:

\noindent \textbf{Start:} Let $\mathcal {P}:=\{P_0\}$. Let $\eta$ be
a totally undefined partial function mapping $\pi$-variables to
bitstrings $[\![$CoSP-terms$]\!]$. Let $\mu$ be a totally undefined
partial function mapping $\pi$-names to bitstrings
$[\![$CoSP-terms$]\!]$. Let $a_1,...,a_n$ denote the free names in
$P_0$. Pick $\{r_i\}_{i=1}^n \in \text{Nonces}_k$ at random
$[\![$Choose a different $r_i\in \textbf{N}_P$$]\!]$. Set $\mu:=\mu
\cup \{a_i:=r_i\}_{i=1}^n$. Send $(r_1,...,r_n)$ to $\mathcal {A}$.

\noindent \textbf{Main loop:} Send $\mathcal {P}$ to $\mathcal {A}$
and expect an evaluation context $E$ from the adversary. Distinguish
the following cases:
\begin{itemize}
\item[$\bullet$]$\mathcal {P}=E[\text{in}(M,x);P_1]$: Request two
bitstrings $[\![$CoSP-terms$]\!]$ $c,m$ from the adversary. If
$c=\text{ceval}_{\eta,\mu}(M)$ $[\![c\cong
\text{eval}^{CoSP}(M\eta\mu)]\!]$, set $\eta:=\eta\cup\{x:=m\}$ and
$\mathcal {P}:=E[P_1]$.

\item[$\bullet$]$\mathcal {P}=E[\nu a;P_1]$: Pick $r\in\text{Nonces}_k$ at random $[\![$ Choose $r\in \textbf{N}_P \backslash \text{range }\mu$$]\!]$, set $\mu:=\mu\cup\{a:=r\}$ and
$\mathcal {P}:=E[P_1]$.

\item[$\bullet$]$\mathcal {P}=E[\text{out}(M_1,N);P_1][\text{in}(M_2,x);P_2]$: If
$\text{ceval}_{\eta,\mu}(M_1)=\text{ceval}_{\eta,\mu}(M_2)\\$
$[\![\text{eval}^{CoSP}(M_1\eta\mu) \cong
\text{eval}^{CoSP}(M_2\eta\mu)]\!]$, set
$\eta:=\eta\cup\{x:=\text{ceval}_{\eta,\mu}(N)\}$
$[\![\eta:=\eta\cup\{x:=\text{eval}^{CoSP}(N\eta\mu)\}]\!]$ and
$\mathcal {P}:=E[P_1][P_2]$.

\item[$\bullet$]$\mathcal {P}=E[\text{let}{\kern 3pt}x=D{\kern 3pt}\text{in}{\kern 3pt}P_1{\kern 3pt} \text{else}{\kern 3pt}P_2]$: If
$m :=\text{ceval}_{\eta,\mu}(D)\neq \bot$ $[\![m:=
\text{eval}^{CoSP}(D\eta\mu)\\\neq\bot]\!]$, set
$\mu:=\mu\cup\{x:=m\}$ and $\mathcal {P}:=E[P_1]$. Otherwise set
$\mathcal {P}:=E[P_2]$

\item[$\bullet$]$\mathcal {P}=E[\text{event}{\kern 3pt}e;P_1]$:
Let $\mathcal {P}:=E[P_1]$ and raise the event $e$.

\item[$\bullet$]$\mathcal {P}=E[!P_1]$: Rename all bound
variables of $P_1$ such that they are pairwise distinct and distinct
from all variables and names in $\mathcal {P}$ and in domains of
$\eta,\mu$, yielding a process $\tilde{P}_1$. Set $\mathcal
{P}:=E[\tilde{P}_1|!P_1]$.

%\item[$\bullet$]$\mathcal {P}=E[P_1|P_2|...]$: Set $\mathcal {P}:=E[\{P_1,P_2,...\}]$.

\item[$\bullet$]$\mathcal {P}=E[\text{out}(M,N);P_1]$: Request a
bitstring $[\![$CoSP-term$]\!]$ $c$ from the adversary. If
$c=\text{ceval}_{\eta,\mu}(M)$ $[\![c\cong
\text{eval}^{CoSP}(M\eta\mu)]\!]$, set $\mathcal {P}:=E[P_1]$ and
send $\text{ceval}_{\eta, \mu}(N)$
$[\![\text{eval}^{CoSP}(N\eta\mu)]\!]$ to the adversary.

\item[$\bullet$]In all other cases, do nothing.
\end{itemize}

We say that a closed process computationally satisfies a $\pi$-trace
property $\wp$ if the list of events raised by its computational
execution is in $\wp$ with overwhelming probability. Then the
theorem in \cite{CoSP} states that for any given computationally
sound implementation of the applied $\pi$-calculus (embedded in the
CoSP model), the symbolic verification of a closed process $P_0$
satisfying a $\pi$-trace property $\wp$ implies $P_0$
computationally satisfies $\wp$.

\section{Computational Soundness Results for SAPIC}

\subsection{SAPIC}

The SAPIC tool was proposed in \cite{SAPIC}. It translates SAPIC
process to multiset rewrite rules, which can be analyzed by the
tamarin-prover \cite{tamarin}. Its language extends the applied
$\pi$ calculus with two kinds of explicit state construsts. The
first kind is functional. It provides the operation for defining,
deleting, retrieving, locking and unlocking the memory states. The
second construct allows to manipulate the global state in the form
of a multiset of ground facts. This state manipulation is similar to
the ``low-level" language of the tamarin-prover and offers a more
flexible way to model stateful protocols. Moreover, the security
property of SAPIC process is expressed by trace formulas. It is
expressive enough to formalize complex properties such as injective
correspondence.

\begin{table}
\begin{center}
  \caption{State constructs of SAPIC calculus}  \label{SAPIC}%
    \begin{tabular}{ll}
\hline
    $\langle P,Q\rangle ::=$ & processes\\
\quad  \quad     ...                            & standard processes \\
\quad \quad    insert $M,N;P    $                & insert\\
 \quad  \quad delete $ M;P     $               & delete\\
 \quad  \quad lookup $M$ as $x$ in $P$ else $Q$  & retrieve\\
 \quad  \quad lock $M;P     $                    & lock \\
  \quad \quad   unlock $M;P     $                 & unlock \\
  \quad \quad $[L]-[e]\rightarrow [R];P$\quad($L,R\in\mathcal{F}^*$)& multiset state construct \\
  \hline
    \end{tabular}%
\end{center}
\end{table}%

\noindent\textbf{Syntax}. We list the two kinds of state constructs
in \textbf{Table \ref{SAPIC}}. Table \ref{standard} and \ref{SAPIC}
together compose the full syntax of SAPIC language. Let
$\Sigma_{fact}$ be a signature that is partitioned into
\emph{linear} and \emph{persistent} fact symbols. We can define the
set of facts as
\[\mathcal {F}:=\{F(M_1,...,M_n)|F\in\Sigma_{fact} \text{ of arity
}n\},
\]

\noindent Given a finite sequence or set of facts $L\in \mathcal
{F}^*$, $lfacts(L)$ denotes the multiset of all linear facts in $L$
and $pfacts(L)$ denotes the set of all persistent facts in $L$.
$\mathcal {G}$ denotes the set of ground facts, i.e., the set of
facts that do not contain variables. Given a set $L$, we denote by
$L^\#$ the set of finite multisets of elements from $L$. We use the
superscript $^\#$ to annotate usual multiset operation, e.g.
$L_1\cup^\#L_2$ denotes the multiset union of multisets $L_1, L_2$.

Note that we do our first restriction in the input construct. In
\cite{SAPIC}, the original SAPIC language allows the input of a term
in the input construct in$(M, N);P$. We use the standard construct
in$(M, x);P$ instead in \textbf{Table \ref{standard}}. We will
explain it later in Section 3.2.

\noindent\textbf{Operational Semantics}. A semantic configuration
for SAPIC calculus is a tuple $(\tilde{n},$ $\mathcal {S},$
$\mathcal {S}^{MS},$ $\mathcal {P},$ $\mathcal {K},$ $\mathcal
{L})$. $\tilde{n}$ is a set of names which have been restricted by
the protocol. $\mathcal {S}$ is a partial function associating the
values to the memory state cells. $\mathcal {S}^{MS}\subseteq
\mathcal {G}^\#$ is a multiset of ground facts. $\mathcal {P} =
\{P_1,...,P_k\}$ is a finite multiset of ground processes
representing the processes to be executed in parallel. $\mathcal
{K}$ is the set of ground terms modeling the messages output to the
environment (adversary). $\mathcal {L}$ is the set of currently
acquired locks. The semantics of the SAPIC is defined by a reduction
relation $\rightarrow$ on semantic configurations. We just list the
semantics of state constructs in \textbf{Fig. \ref{OSSAPIC}}. By
$\mathcal {S}(M)$ we denote $\mathcal {S}(N)$ if $\exists N \in
dom(\mathcal {S}), N =_E M$. By $\mathcal {L}\backslash_E \{M\}$ we
denote $\mathcal {L}\backslash \{N\}$ if $\exists N \in \mathcal
{L}, M =_E N$. The rest are in \cite{SAPIC}.

%Note that the lock and unlock constructs in SAPIC calculus are
%different from those in StatVerif. SAPIC calculus assumes that once
%the state cell $M$ is locked, it can still be operated by the
%parallel processes, but cannot be locked again until a corresponding
%unlock is executed. SAPIC calculus can achieve the same meaning of
%lock semantics by annotating all the processes in parallel with lock
%and unlock constructs. Moreover, there are more functional state
%constructs in SAPIC calculus such as delete (undefine the state
%cells) and the failure branch of read.

\begin{figure*}
{\tiny\begin{align*}
  \left( {\tilde n,\mathcal {S}, \mathcal {S}^{MS},\mathcal {P} \cup^\# \left\{  \text{insert }M,N;P \right\},\mathcal {K},\mathcal {L}} \right) &\xrightarrow{}\left( {\tilde n,\mathcal {S}\cup\{M:= N\}, \mathcal {S}^{MS},\mathcal {P}\cup^\# \left\{P \right\}, \mathcal {K}, \mathcal {L}} \right)  \hfill \\
  \left( {\tilde n,\mathcal {S}, \mathcal {S}^{MS},\mathcal {P} \cup^\# \left\{  \text{delete }M;P \right\},\mathcal {K},\mathcal {L}} \right) &\xrightarrow{}\left( {\tilde n,\mathcal {S}\cup\{M:= \bot\}, \mathcal {S}^{MS},\mathcal {P}\cup^\# \left\{P \right\}, \mathcal {K}, \mathcal {L}} \right)  \hfill \\
  \left( {\tilde n,\mathcal {S}, \mathcal {S}^{MS},\mathcal {P} \cup^\# \left\{  \text{lookup }M\text{ as }x\text{ in }P\text{ else }Q \right\},\mathcal {K},\mathcal {L}} \right)&\xrightarrow{}\left( {\tilde n,\mathcal {S}, \mathcal {S}^{MS},\mathcal {P}\cup^\# \left\{P\{V/x\} \right\}, \mathcal {K}, \mathcal {L}} \right)\text{if }\mathcal {S}(M)=_E V  \\
  \left( {\tilde n,\mathcal {S}, \mathcal {S}^{MS},\mathcal {P} \cup^\# \left\{  \text{lookup }M\text{ as }x\text{ in }P\text{ else }Q \right\},\mathcal {K},\mathcal {L}} \right) &\xrightarrow{}\left( {\tilde n,\mathcal {S}, \mathcal {S}^{MS},\mathcal {P}\cup^\# \left\{Q\} \right\}, \mathcal {K}, \mathcal {L}} \right) \text{if }\mathcal {S}(M)=\bot \hfill \\
  \left( {\tilde n,\mathcal {S}, \mathcal {S}^{MS},\mathcal {P} \cup^\# \left\{  \text{lock }M;P \right\},\mathcal {K},\mathcal {L}} \right) &\xrightarrow{}\left( {\tilde n,\mathcal {S}, \mathcal {S}^{MS},\mathcal {P}\cup^\# \left\{P \right\}, \mathcal {K}, \mathcal {L}\cup\{M\}} \right) \text{if }M\notin_E \mathcal {L} \hfill \\
  \left( {\tilde n,\mathcal {S}, \mathcal {S}^{MS},\mathcal {P} \cup^\# \left\{  \text{unlock }M;P \right\},\mathcal {K},\mathcal {L}} \right) &\xrightarrow{}\left( {\tilde n,\mathcal {S}, \mathcal {S}^{MS},\mathcal {P}\cup^\# \left\{P \right\}, \mathcal {K}, \mathcal {L}\backslash_E\{M\}} \right) \text{if }M\in_E \mathcal {L} \hfill \\
  \left( {\tilde n,\mathcal {S}, \mathcal {S}^{MS},\mathcal {P} \cup^\# \left\{  [L]-[e]\rightarrow[R];P \right\},\mathcal {K},\mathcal {L}} \right) &\xrightarrow{e}\left( {\tilde n,\mathcal {S}, \mathcal {S}^{MS}\backslash lfacts(L')\cup^\# R',\mathcal {P}\cup^\# \left\{P\tau \right\}, \mathcal {K}, \mathcal {L}} \right) \hfill \\
  \text{if }\exists \tau,L',R'. {\kern 4pt} \tau \text{ grounding for }L,R \text{ such that }L'=_E L\tau,&R'=_E R\tau, \text{ and } lfacts(L')\subseteq^\# \mathcal {S}^{MS}, pfacts(L')\subset \mathcal {S}^{MS} \hfill \\
  \end{align*}}
\caption{The semantics of SAPIC}\label{OSSAPIC}
\end{figure*}

\noindent\textbf{Security Property}. With the operational semantics,
we can give out the definition of SAPIC trace property. The set of
traces of a closed SAPIC process $P$, written $traces(P)$, defines
all its possible executions. In SAPIC, security properties are
described in a two-sorted first-order logic, defined as the trace
formula. Given a closed SAPIC process $P$, a trace formula $\phi$ is
said to be \emph{valid} for $P$, written $traces(P)\vDash^{\forall}
\phi$, if all the traces of $P$ satisfies $\phi$. $\phi$ is said to
be \emph{satisfiable} for $P$, written $traces(P)\vDash^{\exists}
\phi$, if there exists a trace of $P$ satisfies $\phi$. Note that
$traces(P)\vDash^{\exists} \phi$ \emph{iff}
$traces(P)\nvDash^{\forall} \neg\phi$. It means the verification of
satisfiability can be transformed to the falsification of validity.
Thus in the following, we only consider the validity of trace
formula. We can transform its definition to trace property in the
sense of Definition 2 by requiring that
$\wp:=\left\{tr|tr\vDash\phi\right\}$. Then we get the following
definition of SAPIC trace property.

\noindent\textbf{Definition 7 (SAPIC Trace Property)}. Given a
closed SAPIC process $P$, we define the set of traces of $P$ as
\begin{align*}
traces(P)=\{ [e_1,...,e_m]|& (\emptyset,\emptyset,\emptyset,\{P\},fn(P),\emptyset)\xrightarrow{}^*\xrightarrow{e_1}(\tilde n_1,\mathcal {S}_1, \mathcal{S}^{MS}_1,\mathcal{P}_1,\mathcal {K}_1,\mathcal{L}_1)\\
&\xrightarrow{}^*\xrightarrow{e_2}\cdots\xrightarrow{}^*\xrightarrow{e_m}(\tilde n_m,\mathcal {S}_m, \mathcal{S}^{MS}_m,\mathcal{P}_m,\mathcal {K}_m,\mathcal{L}_m)\}\\
\end{align*}

\noindent A SAPIC trace property $\wp$ is an efficiently decidable
and prefix-closed set of strings. A process $P$ symbolically
satisfies the SAPIC trace property $\wp $ if we have
$traces(P)\subseteq \wp $.

\subsection{CS Results of the Calculus}.

SAPIC language only has semantics in the symbolic model. We need to
introduce the computational execution model of SAPIC process. It is
not a trivial extension of the computational execution model of the
applied $\pi$ calculus in Definition 5. We first restrict the
pattern matching in the original SAPIC input construct because for
some cases, it cannot be performed by any sound computational model.
Then we set up the computational execution model for the two kinds
of global states in SAPIC. Note that the CoSP framework does not
immediately support nodes for the operation of functional states and
multiset states. We will encode them into the CoSP protocol node
identifiers and mechanize the two kinds of state constructs by using
CoSP protocol tree.

First, we need to explain the restriction of the input construct.
Note that we use the standard syntax of applied $\pi$ calculus as
part of the syntax of SAPIC language in \textbf{Table \ref{SAPIC}}.
In \cite{SAPIC}, the original SAPIC process allows the input of a
term in the input construct in$(M, N);P$ where it receives a ground
term $N'$ on the channel $M$, does a pattern matching to find a
substitution $\tau$ such that $N'=_E N\tau$, and then proceeds by
$P\tau$. However, we find that it is impossible to embed it into the
CoSP framework. As in Definition 5, the computational execution of
the calculus receives the bitstring $m$ from the adversary. Then the
interactive machine $Exec_{P_0}(1^k)$ should extract from $m$ the
sub-bitstrings corresponding to the subterms in the range of $\tau$.
This is impossible for some cases. One example is the input process
$P:=\text{in}(c, h(x))$ where the adversary may generate a name $t$,
compute and output the term $h(t)$ on the channel $c$. It has no
computational execution model since the protocol does not know how
to bind the variable $x$ ($h(\cdot)$ is not invertible). Thus in the
following, we do our \emph{first restriction} that the SAPIC input
construct should be in the form in($M, x$).

Then we show how to embed the two kinds of states into the CoSP
framework and mechanize the state constructs.
Our computational execution model maintains a standard protocol
state that consists of the current process $\mathcal {P}$, an
environment $\eta$, and an interpretation $\mu$ as in Definition 5. Moreover, we extend
the protocol state with a set $S$ including all the pairs $(M,N)$ of
the functional state cells $M$ and their associated values $N$, a
set $\Lambda$ of all the currently locked state cells, and a
multiset $S^{MS}$ of the current ground facts. We denote by
$dom(S):=\{m|(m,n)\in S\}$ the set of state cells in $S$ ($S$ can be
seen as a partial function and $dom(S)$ is its domain). In each step
of the execution, the adversary receives the process $\mathcal {P}$
and sends back an evaluation context $E$ where $\mathcal {P} =
E[\mathcal {P}_1]$ to schedule the proceeding to $\mathcal {P}_1$.
In addition to the standard cases operated in Definition 5, we need
to mechanize the functional and multiset state constructs according
to the protocol states $S$, $\Lambda$, and $S^{MS}$. We implement
the procedures as CoSP sub-protocols. Note that our encoding
should keep the efficiency of the resulting CoSP protocol and cannot
introduce an unacceptable time cost for computational execution. In
the following, we respectively explain how to embed the two kinds of
state constructs.

\noindent\textbf{Embedding the functional state}. For the functional
state constructs in SAPIC, the state cells and their associated
values are $\pi$-terms. If we encode them directly as $\pi$-terms in
the set $S$, its size would grow exponentially, and the resulting
CoSP protocol is not efficient. To solve this problem, we store the
state cell $M$ and its value $N$ as CoSP-terms in the sets $S$ and
$\Lambda$. The CoSP-terms can be encoded by the indexes of the nodes
in which they were created (or received). In this setting, the
CoSP-terms are treated as black boxes by the CoSP protocol with a
linear size.

\begin{figure}
\begin{center}
  \includegraphics[width=8cm]{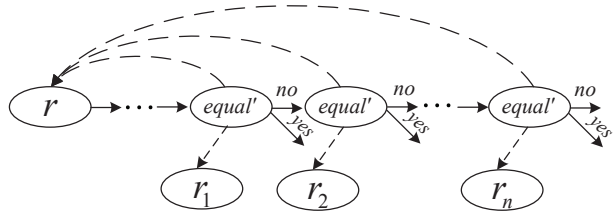}
  \caption{Sub-protocol $f_{mem}$ for decision of set-membership}\label{CoSP4MS}
\end{center}
\end{figure}

However, we have to pay extra cost for this setting. For a finite
set of CoSP terms, such as $dom(S)$ or $\Lambda$, we need to
formalize the decision of set-membership. It can be done with the
help of \emph{parameterized CoSP protocols}, which act as
sub-protocols with formal parameters of CoSP nodes and can be
plugged into another CoSP protocol tree. Its definition is
introduced in \cite{CS14}. We denote by $f_{\text{mem}}$ the
decision of set-membership relation: if $\exists r_i \in \Lambda,
r_i\cong r$, where $r$ is a CoSP-term, $\Lambda=\{r_1,...,r_n\}$ is
a set of CoSP-terms. It can be accomplished by a sequence of $n$
CoSP computation nodes for the destructor $equal'$ as in
\textbf{Fig. \ref{CoSP4MS}}. The success-edge of
$f_{\text{mem}}(\Lambda;r)$ corresponds to each \texttt{yes}-edge.
The failure-edge corresponds to the \texttt{no}-edge of the last
computation node. With this sub-protocol, we can embed the
functional state constructs in the execution model of SAPIC. The
computation steps of the embedding would not grow exponentially.
Decision of set-membership costs no more than the size of the set,
which is bounded by the reduction steps $t$. Thus there exists a
polynomial $p$, such that the computation steps of embedding is
bounded by $p(t)$.

\noindent\textbf{Embedding the multiset state}. For the multiset
state, we keep a multiset $S^{MS}$ of the current ground facts.
In the execution model, we need to encode the multiset state
construct $[L]-[e]\rightarrow [R];P$ by using CoSP sub-protocol
$f_{match}$. As in \textbf{Fig. \ref{OSSAPIC}}, the
SAPIC process tries to match each fact in the sequence $L$ to the
ground facts in $S^{MS}$ and, if successful, adds the corresponding
instance of facts $R$ to $S^{MS}$. We denote by $fv(L)$ the
set of variables in $L$ that are not under the scope of a previous
binder. The variables $x \in fv(L)$ should be bound by the pattern matching.
For the reason of efficiency, we store the arguments of
ground facts in $S^{MS}$ as CoSP-terms rather than
$\pi$-terms\footnote{Otherwise, the length of $\pi$-terms may grow
exponentially by the iterated binding of variables. One example is
the construct $!([\text{Iter}(x)]-[]\rightarrow
[\text{Iter}(fun(x,x))]$).}, as we have done in the case of
functional state. $S^{MS}$ can only be altered using the multiset
state construct $[L]-[e]\rightarrow [R];P$. Given a closed SAPIC
process, the maximum length of $R$ (counted by the number of
fact symbols in $R$) is a constant value. In each
execution step, the multiset state construct can proceed at most
once. Thus the size of $S^{MS}$ is bounded by a polynomial in the
number of execution steps (taken CoSP-terms as blackboxes).

%Note that the pattern matching in this construct is different from
%the one in the input construct. In this case, only the SAPIC process
%can operate the multiset and do the pattern matching. This provides
%a possibility to handle it. We will explain it later.

When designing the sub-protocol $f_{match}$ for the multiset
state construct, we should solve the pattern matching
problem, which is similar to the previous one in the input construct.
To solve this problem, we need to do our \emph{second restriction}.
In the multiset state construct $[L]-[e]\rightarrow [R];P$, we
require that: (i) it is well-formed (Definition 12 in \cite{SAPIC});
(ii) $\forall F(M_1, ..., M_n)\in L$, either $M_i\in fv(L)$ or
$fv(M_i)=\emptyset$ for all $1\leq i\leq n$. It means that the free
variables of $L$ can only occur as the arguments of the facts in
$L$. By (i), the well-formed requirement, we have $fv(R) \subseteq
fv(L)$. Thus all the facts added into the current multiset state
$S^{MS}$ are ground. By (ii), we can match each variable
in $fv(L)$ to the corresponding arguments of the ground facts in
$S^{MS}$ and find the substitution $\tau$ for $fv(L)$ in the
execution. Note that our second restriction is necessary for the CS
results. Otherwise, if we allow the free variables in $fv(L)$ occur
as the subterms of the arguments of facts, it might lead to a
mismatch case as we have described in the input construct.

The second restriction does not make the multiset state construct useless.
All the examples in \cite{SAPIC} using this construct meet our requirements.
Moreover, this style of state manipulation is the underlying
specification language of the tamarin tool \cite{tamarin}. Even
considering our restriction, the tamarin tool is still useful to
model security protocols. The example is the NAXOS protocol for the
eCK model formalized in \cite{tamarin}.

In the following, we will give out the sub-protocol $f_{match}$ of
the pattern matching. Since $f_{match}$ is plugged
in the execution model of SAPIC, it assumes an initial protocol state which
includes an environment $\eta$, an interpretation $\mu$, and a
multiset $S^{MS}$ of the current ground facts. For each multiset
state construct $[L]-[e]\rightarrow [R]$, $f_{match}$ tries to find
a substitution $\tau'$ from $fv(L)$ to CoSP-terms, such that
$lfacts(L)\eta'\mu \subseteq^\# S^{MS}$ and $ pfacts(L)\eta'\mu
\subset S^{MS}$, where $\eta' = \eta \cup \tau'$. For simplicity, we
denote by $f/(n,k)$ a $\pi$-fact such that $f/(n,k) =
F(M_1,...,M_k)\in\mathcal {F}$ and $\{M_i\}^{k}_{i=1}$ are
$\pi$-terms including $n$ variables. A $\pi$-fact $f/(0,k)$ is
ground.

\noindent\textbf{Definition 8 (Sub-protocol of Pattern Matching)}.
Let $\eta$ be a partial function mapping variables to CoSP-terms,
let $\mu$ be a partial function mapping $\pi$-names to CoSP-terms,
let $S^{MS}$ be a multiset of facts whose arguments are CoSP-terms.
Let $[L]-[e]\rightarrow [R];P$ be a multiset state construct with
our restriction. We define the sub-protocol $f_{match}$ which
contains two stages respectively for the pattern matching of linear
and persistent facts in $L$:

\noindent \textbf{Start}. For stage 1, let $\tau'$ be a totally
undefined partial function mapping variables to CoSP-terms. Set
$S_{rest}:=S^{MS}$. Let $L_{rest}:=lfacts(L)$ and
$L_{linear}:=\emptyset$ be two multisets of $\pi$-facts.

\noindent \textbf{Loop}. Choose a $\pi$-fact $l/(n,k) \in^\#
L_{rest}$, match it to all the fact $f \in^\# S_{rest}$ with the
same fact symbol by performing the following steps i)-iii).
If any check in step ii) is failed, choose the next $f \in^\# S_{rest}$ to
match. If there is no matching with $l/(n,k)$ for any facts in $S_{rest}$,
stop and go to the failure-edge.
\begin{itemize}
\item[i)] For $n$ variables $x_i$ in $l/(n,k)$,
pick up $x_i \notin dom(\eta)\cup dom(\tau')$(i.e., the free
variables in $l$), set $\tau'':=\tau'\cup\{x_i\mapsto s_i| 1\leq i
\leq n, x_i \notin dom(\eta)\cup dom(\tau')\}$ by mapping $x_i$ to
the CoSP-term $s_i$ with the same position in $f$. This can be done
since we require free variables should be the arguments of facts.

\item[ii)] For $k$ arguments of $l/(n,k) = F(M_1,...,M_k)$,
use the CoSP computation node to check whether $t_j \cong
\text{eval}^{CoSP}(M_j\eta'\mu)$ for $j = 1, ..., k$, where $t_j$ is
the argument of $f$ with the same position, $\eta'=\eta\cup \tau''$.
This can be done since $dom(\eta)\cap dom(\tau'') = \emptyset$.

\item[iii)] If all the checks in step ii) pass, we set $L_{rest}:=L_{rest}\backslash^\# \{l/(n,k)\}$,
$S_{rest}:=S_{rest}\backslash^\# \{f\}$, $L_{linear}:=L_{linear}
\cup^\# \{l/(n,k)\}$, and $\tau' = \tau''$. Loop while
$L_{rest}\neq\emptyset$.
\end{itemize}

\noindent Stage 2 is similar. We perform the above algorithm of
stage 1 without $^\#$. In the Start, let $\tau'$ be the one we have
achieved in stage 1, set $L_{rest}:=pfacts(L)$, $S_{rest}:=S^{MS}$,
and do not change $S_{rest}$ in step iii) of the Loop. If both the
two stages are successful, $f_{match}$ goes to the success-edge.

All the steps in $f_{match}$ can be performed by CoSP nodes. By the
conditions in step ii), if successful, $f_{match}$ will find $\tau'
$ and $\eta' = \eta \cup \tau'$ such that $ lfacts(L)\eta'\mu
\subseteq^\# S^{MS}$ and $pfacts(L)\eta'\mu \subset S^{MS}$. Thus we
encode the pattern matching of multiset state construct into the CoSP
sub-protocol $f_{match}$.

Then we need to explain that the embedding way does not cost
unacceptably high. The time complexity of the above sub-protocol
(measured by the CoSP nodes) is approximately the size of $S^{MS}$
times the size of $L$. Given a closed SAPIC process, the
maximum size of $L$ is a constant number and the size of $S^{MS}$ is
polynomial in the execution steps $t$. Thus there exists a
polynomial $p$, such that the computation steps of encoding is
bounded by $p(t)$.

Now we could give out the definition of computational execution
model of SAPIC in Definition 9. It is an interactive machine
$Exec_{P_0}^S(1^k)$ that executes the SAPIC process and communicates
with a probabilistic polynomial-time adversary. The model maintains
a protocol state as 6-tuple $(\mathcal {P}, \eta, \mu, S, \Lambda,
S^{MS})$. The definition of the evaluation context is similar to
that of the applied $\pi$ calculus. We write $E[P]=\mathcal {P}\cup
\{P\}$.

%For a partial function $g$, we define the function $f:=g\cup
%\{a\mapsto b\}$ with $dom(f)=dom(g)\cup\{a\}$ as $f(a):=b$ and
%$f(x):=g(x)$ for $x\neq a$.

In order to relate the symbolic and the computational semantics of a
SAPIC process, we also define an additional symbolic execution for
closed SAPIC processes as a technical tool as in \cite{CoSP}. It is
a direct analogue of the computational execution model and denoted
by $SExec_{P_0}^S$. The difference between $Exec_{P_0}^S(1^k)$ and
$SExec_{P_0}^S$ is that the latter one operates on CoSP-terms rather than
bitstrings: It computes CoSP-terms $M\eta\mu$ and
$eval^{CoSP}D\eta\mu$ instead of bitstrings ceval$_{\eta,\mu}(M)$
and ceval$_{\eta, \mu}(D)$, it compares the CoSP-terms using
CoSP-destructor $\cong$ instead of checking for equality of
bitstrings, and it chooses a fresh nonce $r\in\textbf{N}_P$ instead
of choosing a random bitstring $r$ as value for a new protocol name.

Due to the limited space, we merge the Definition 10 of the symbolic
execution of SAPIC into the Definition 9 of the computational one.
It is marked by $[\![...]\!]$. In the main loop, we only present the
cases of SAPIC state constructs. For the standard cases, the
execution model performs in the same way as the applied $\pi$
calculus model does.

\noindent\textbf{Definition 9 $[\![$10$]\!]$ (Computational
$[\![$Symbolic$]\!]$ Execution of SAPIC)}.  Let $P_0$ be a closed
SAPIC process (where all bound variables and names are renamed such
that they are pairwise distinct and distinct from all unbound ones).
Let $\mathcal {A}$ be an interactive machine called the adversary.
We define the computational $[\![$symbolic$]\!]$ execution of SAPIC
calculus as an interactive machine $Exec_{P_0}^S(1^k)$ that takes a
security parameter $k$ as argument $[\![$interactive machine
$SExec_{P_0}^S$ that takes no argument$]\!]$ and interacts with
$\mathcal {A}$:

\noindent \textbf{Start:} Let $\mathcal {P}:=\{P_0\}$. Let $\eta$ be
a totally undefined partial function mapping $\pi$-variables to
bitstrings $[\![$CoSP-terms$]\!]$, let $\mu$ be a totally undefined
partial function mapping $\pi$-names to bitstrings
$[\![$CoSP-terms$]\!]$, let $S$ be an initially empty set of pairs
of bitstrings $[\![$CoSP-terms$]\!]$. Let $S^{MS}$ be an initially
empty multiset of facts whose arguments are bitstrings
$[\![$CoSP-terms$]\!]$. Let $\Lambda$ be an initially empty set of
bitstrings $[\![$CoSP-terms$]\!]$. Let $a_1,...,a_n$ denote the free
names in $P_0$. Pick $\{r_i\}_{i=1}^n \in \text{Nonces}_k$ at random
$[\![$Choose a different $r_i\in \textbf{N}_P$$]\!]$. Set $\mu:=\mu
\cup \{a_i:=r_i\}_{i=1}^n$. Send $(r_1,...,r_n)$ to $\mathcal {A}$.

\noindent \textbf{Main loop:} Send $\mathcal {P}$ to $\mathcal {A}$
and expect an evaluation context $E$ from the adversary. Distinguish
the following cases:
\begin{itemize}
\item[$\bullet$]For the standard cases, the execution model performs
the same way as in Definition 5 $[\![$6$]\!]$.

\item[$\bullet$]$\mathcal {P} = E[\text{insert }M,N;P_1]$:
Set $m:= \text{ceval}_{\eta, \mu}(M), n:= \text{ceval}_{\eta,
\mu}(N)$ $[\![m:=\text{eval}^{CoSP}(M\eta\mu),
n:=\text{eval}^{CoSP}(N\eta\mu)]\!]$. Plug in $f_{mem}$ to decide if
$\exists (r',r)\in S, r'=m$ $[\![r'\cong m]\!]$. For the
success-edge, set $\mathcal {P}:=E[P_1]$ and $S:=S\backslash
\{(r',r)\}\cup\{(m,n)\}$. For the failure-edge, set
$\mathcal{P}:=E[P_1]$ and $S:=S\cup\{(m,n)\}$.

\item[$\bullet$]$\mathcal {P} = E[\text{delete }M;P_1]$:
Set $m:= \text{ceval}_{\eta, \mu}(M)$
$[\![m:=\text{eval}^{CoSP}(M\eta\mu)]\!]$. Plug in $f_{mem}$ to
decide if $\exists (r',r)\in S, r' = m$ $[\![r'\cong m]\!]$. For the
success-edge, set $\mathcal {P}:=E[P_1]$ and
$S:=S\backslash\{(r',r)\}$. For the failure-edge, set
$\mathcal{P}:=E[P_1]$.

\item[$\bullet$]$\mathcal {P} = E[\text{lookup }M\text{ as }x\text{ in }P_1\text{ else }P_2]$:
Set $m:= \text{ceval}_{\eta, \mu}(M)$
$[\![m:=\text{eval}^{CoSP}\\(M\eta\mu)]\!]$. Plug in $f_{mem}$ to
decide if $\exists (r',r)\in S,r'=m$ $[\![r'\cong m]\!]$. For the
success-edge, set $\mathcal {P}:=E[P_1]$ and
$\eta:=\eta\cup\{x:=r\}$. For the failure-edge, set
$\mathcal{P}:=E[P_2]$.

\item[$\bullet$]$\mathcal {P} = E[\text{lock }M;P_1]$:
Set $m:= \text{ceval}_{\eta, \mu}(M)$
$[\![m:=\text{eval}^{CoSP}(M\eta\mu)]\!]$. Plug in $f_{mem}$ to
decide if $\exists r'\in \Lambda,r'=m$ $[\![r'\cong m]\!]$. For the
success-edge, do nothing. For the failure-edge, set $\mathcal
{P}:=E[P_1]$ and $\Lambda:=\Lambda\cup\{ m\}$.

\item[$\bullet$]$\mathcal {P} = E[\text{unlock }M;P_1]$:
Set $m:= \text{ceval}_{\eta, \mu}(M)$
$[\![m:=\text{eval}^{CoSP}(M\eta\mu)]\!]$. Plug in $f_{mem}$ to
decide if $\exists r'\in \Lambda,r'=m$ $[\![r'\cong m]\!]$. For the
success-edge, set $\mathcal {P}:=E[P_1]$ and
$\Lambda:=\Lambda\backslash\{ r' \}$. For the failure-edge, do
nothing.

\item[$\bullet$]$\mathcal {P} = E[[L]-[e]\rightarrow[R];P_1]$: Plug in $f_{match}$ to find a substitution
$\tau'$ from $fv(L)$ to bitstrings $[\![$CoSP-terms$]\!]$, such that
$lfacts(L)\eta'\mu \subseteq^\# S^{MS}$ and $ pfacts(L)\eta'\mu
\subset S^{MS}$, where $\eta' = \eta \cup \tau'$. For the
success-edge, set $\mathcal {P}:=E[P_1]$,
$S^{MS}:=S^{MS}\backslash^\# lfacts(L)\eta'\mu \cup R\eta'\mu$,
$\eta:=\eta'$, and raise the event $e$. For the failure-edge, do
nothing.

\item[$\bullet$]In all other cases, do nothing.
\end{itemize}

For a given polynomial-time interactive machine $\mathcal {A}$, a
closed SAPIC process $P_0$, and a polynomial $p$, let
$Events_{\mathcal {A},P_0,p}^S(k)$ be the distribution for the list
of events raised within the first $p(k)$ computational steps
(jointly counted for $\mathcal {A}(1^k)$ and $Exec_{P_0}^S(1^k)$).
Then the computational fulfillment of SAPIC trace properties can be
defined as follows.

\noindent\textbf{Definition 11 (Computational SAPIC Trace
Properties)}. Let $P_0$ be a closed process, and $p$ a polynomial.
We say that $P_0$ computationally satisfies a SAPIC trace property
$\wp $ if for all polynomial-time interactive machines $\mathcal
{A}$ and all polynomials $p$, we have that
$\text{Pr}[Events_{\mathcal {A},P_0,p}^S(k)\in \wp ]$  is
overwhelming in $k$.

%In order to relate the symbolic and the computational semantics of a
%SAPIC process, we define an additional symbolic execution for closed
%SAPIC processes as a technical tool as \cite{CoSP} did. It is a
%direct analogue of the computational semantics presented in
%Definition 9. The symbolic execution model is defined by an
%interactive machine $SExec_{P_0}^S$ in Definition 11.
%
%The difference between the Definition 9 and 11 is that the latter
%one operates on CoSP-terms rather bitstrings: It computes CoSP-terms
%$M\eta\mu$ and $eval^{CoSP}D\eta\mu$ instead of bitstrings
%ceval$_{\eta,\mu}(M)$ and ceval$_{\eta, \mu}(D)$. It compares the
%CoSP-terms using CoSP-destructor $\cong$ instead of checking for
%equality of bitstrings, and it chooses a fresh nonce
%$r\in\textbf{N}_P$ instead of choosing a random bitstring $r$ as
%value for a new protocol name.

Then we should explain that $SExec_{P_0}^S$ can be realized by a
CoSP protocol tree. The state of the machine $SExec_{P_0}^S$
includes a tuple $(\mathcal {P}, \mu, \eta, S, S^{MS}, \Lambda)$. It
is used as a node identifier. CoSP-terms should be encoded by the
indexes in the path from the root to the node in which they were
created (or received). The process $\mathcal {P}$, the fact symbols
in $S^{MS}$, and the $\pi$-names in $dom(\mu)$ will be encoded as
bitstrings. We plug two sub-protocols, $f_{mem}$ and $f_{match}$,
into the CoSP protocol respectively for the decision of set-membership
in the functional state constructs, and for the pattern
matching in the multiset state constructs. We have explained that
these two sub-protocols do not introduce an unacceptable cost. The
operation of raising event $e$ can be realized using a control node
with one successor that sends $(event,e)$ to the adversary. Given a
sequence of nodes $\underline{\nu}$, we denote by
$events(\underline{\nu})$ the events $\underline{e}$ raised by the
event nodes in $\underline{\nu}$. We call this resulting CoSP
protocol $\Pi_{P_0}^S$.

\noindent\textbf{Definition 12}. $SExec_{P_0}^S$ satisfies a SAPIC
trace property $\wp $ if in a finite interaction with any Dolev-Yao
adversary, the sequence of events raised by $SExec_{P_0}^S$ is
contained in $\wp $.

Before we prove Theorem 1 of the CS result of SAPIC, we first state
and prove three lemmas. Lemma 1 relates the computational/symbolic
execution of SAPIC calculus and the CoSP protocol $\Pi_{P_0}^S$.
Lemma 2 states that $\Pi_{P_0}^S$ is efficient. Lemma 3 asserts that
the symbolic execution is a safe approximation for SAPIC. Theorem 1
states that the computationally sound implementation of the symbolic
model of applied $\pi$ calculus implies the CS result of SAPIC
calculus. We present the proofs in Appendix A.

\noindent\textbf{Lemma 1}. $SExec_{P_0}^S$ satisfies a trace
property $\wp $ \emph{iff} $\Pi_{P_0}^S$ satisfies $events^{-1}(\wp
)$. Moreover, $P_0$ computationally satisfies $\wp $ \emph{iff}
$(\Pi_{P_0}^S, A)$ computationally satisfies $events^{-1}(\wp )$.
Both are in the sense of Definition 2.

\noindent\textbf{Lemma 2}. The CoSP protocol $\Pi_{P_0}^S$ is
efficient.

\noindent\textbf{Lemma 3}. If a SAPIC
closed process $P_0$ symbolically satisfies a SAPIC trace property
$\wp $ in the sense of Definition 7, then $SExec_{P_0}^S$ satisfies
$\wp $ in the sense of Definition 12.

\noindent\textbf{Theorem 1 (CS in SAPIC)}. Assume that the
computational implementation of the applied $\pi$ calculus is a
computationally sound implementation (in the sense of Definition 3)
of the symbolic model of applied $\pi$ calculus (Definition 4) for a
class \textbf{P} of protocols. If a closed SAPIC process $P_0$
symbolically satisfies a SAPIC trace property $\wp$ (Definition 7),
and $\Pi_{P_0}^S\in\textbf{P}$, then $P_0$ computationally satisfies
$\wp$ (Definition 11).

\section{Computational Soundness Result for StatVerif}

StatVerif was proposed in \cite{Statverif}. Its process language is
an extension of the ProVerif process calculus with only functional
state constructs. StatVerif is limited to the verification of
secrecy property.

In this section, we first encode the StatVerif processes into a
subset of SAPIC processes. Then we prove that our encoding is
able to capture secrecy of stateful protocols by using SAPIC trace
properties. Finally with the CS result of SAPIC, we can directly
obtain the CS result for StatVerif calculus. Note that our encoding
way shows the differences between the semantics of state constructs
in these two calculi.

\begin{table}
\begin{center}
  \caption{State constructs of StatVerif calculus}  \label{Statverif}%
    \begin{tabular}{ll}
\hline
    $\langle P,Q\rangle ::=$ & processes\\
\quad  \quad     ...                             & standard processes \\
\quad \quad    $[s\mapsto M]$                    & initialize\\
 \quad  \quad $s:=M;P$                           & assign\\
 \quad  \quad read $s$ as $x;P$                  & read\\
 \quad  \quad lock; $P$                          & lock state\\
  \quad \quad unlock; $P$                        & unlock state\\
  \hline
    \end{tabular}%
\end{center}
\end{table}%

\noindent\textbf{Syntax}. We first review the StatVerif calculus proposed in
\cite{Statverif}. We list the explicit functional state constructs
in \textbf{Table \ref{Statverif}}. Table \ref{standard} and
\ref{Statverif} together compose the full syntax of StatVerif
calculus. Note that the state constructs are subject to the following two
additional restrictions:
\begin{itemize}
\item[$\bullet$] $[s\mapsto M]$ may occur only once for a given cell
name $s$, and may occur only within the scope of name restriction, a
parallel and a replication.

\item[$\bullet$] For every lock$;P$, the part $P$ of the process
must not include parallel or replication unless it is after an
unlock construct.
\end{itemize}

\noindent\textbf{Operational Semantics}. A semantic configuration
for StatVerif is a tuple $(\tilde{n}, \mathcal {S}, \mathcal {P},
\mathcal {K})$. $\tilde{n}$ is a finite set of names. $\mathcal {S}
= \{s_i := M_i\}$ is a partial function mapping cell names $s_i$ to
their associated values $M_i$. $\mathcal {P} =
\{(P_1,\beta_1),...,(P_k,\beta_k)\}$ is a finite multiset of pairs
where $P_i$ is a process and $\beta_i \in \{0,1\}$ is a boolean
indicating whether $P_i$ has locked the state. For any $1 \leq i
\leq k$, we have at most one $\beta_i = 1$. $\mathcal {K}$ is a set
of ground terms modeling the messages output to the environment
(adversary). The semantics of StatVerif calculus is defined by
transition rules on semantic configurations. We do a little change
to the original semantics by adding two labelled transitions for the
input and output of adversary. With these rules, we can define
secrecy property without explicitly considering the adversary
processes. We list these two rules and the semantics of state
constructs in \textbf{Fig. \ref{OSStat}}. The rest are in
\cite{Statverif}.

\begin{figure*}
{\tiny\begin{align*}
  \left( {\tilde n,\mathcal S,\mathcal {P}  \cup \left\{ {\left( {[s \mapsto M],0 } \right)} \right\}, \mathcal {K}} \right) &\xrightarrow{{}}\left( {\tilde n,\mathcal S \cup \{ s := M\} ,\mathcal {P}, \mathcal {K}} \right) {\text{ if }} s \in \tilde n{\text{ and }} s \notin dom(\mathcal {S}) \hfill \\
  \left( {\tilde n,\mathcal S ,\mathcal {P}  \cup \left\{ {\left( {s: = N;P,\beta} \right)} \right\}, \mathcal {K}} \right) &\xrightarrow{{}}\left( {\tilde n,\mathcal S\cup \{s := N\} ,\mathcal {P}  \cup \left\{ {\left( {P,\beta} \right)} \right\}, \mathcal {K}} \right) {\text{ if }} s \in dom(\mathcal {S}) {\text{ and }} \forall (Q,\beta')\in \mathcal {P},\beta'=0 \hfill \\
  \left( {\tilde n,\mathcal S ,\mathcal {P}  \cup \left\{ {\left( {\text{read } s \text{ as } x;P,\beta} \right)} \right\}, \mathcal {K}} \right) &\xrightarrow{{}}\left( {\tilde n,\mathcal {S},\mathcal {P}  \cup \left\{ {\left( {P\{ \mathcal {S}(s)/x\} ,\beta} \right)} \right\}, \mathcal {K}} \right) {\text{ if }} s \in dom(\mathcal {S}) {\text{ and }}  \forall (Q,\beta')\in \mathcal {P},\beta'=0 \hfill \\
  \left( {\tilde n,\mathcal S ,\mathcal {P}  \cup \left\{ {\left( {\text{lock};P,0} \right)} \right\}, \mathcal {K}} \right) &\xrightarrow{{}}\left( {\tilde n,\mathcal {S},\mathcal {P}  \cup \left\{ {\left( {P,1 } \right)} \right\}, \mathcal {K}} \right) {\text{ if }} \forall (Q,\beta')\in \mathcal {P},\beta'=0 \hfill \\
  \left( {\tilde n,\mathcal S ,\mathcal {P}  \cup \left\{ {\left( {\text{unlock};P,1} \right)} \right\}, \mathcal {K}} \right) &\xrightarrow{{}}\left( {\tilde n,\mathcal {S},\mathcal {P}  \cup \left\{ {\left( {P,0 } \right)} \right\}, \mathcal {K}} \right) \hfill \\
  \left( {\tilde n,\mathcal S ,\mathcal {P}  \cup \left\{ {\left( {\text{out}(M,N);P,\beta} \right)} \right\}, \mathcal {K}} \right) &\xrightarrow{{K(N)}}\left( {\tilde n,\mathcal {S},\mathcal {P}  \cup \left\{ {\left( {P,\beta } \right)} \right\}, \mathcal {K}\cup\{N\}} \right) {\text{ if }} \nu \tilde{n}.\mathcal {K} \vdash M \hfill \\
  \left( {\tilde n,\mathcal S ,\mathcal {P}  \cup \left\{ {\left( {\text{in}(M,x);P,\beta} \right)} \right\}, \mathcal {K}} \right) &\xrightarrow{{K(M,N)}}\left( {\tilde n,\mathcal {S},\mathcal {P}  \cup \left\{ {\left( {P\{N/x\},\beta } \right)} \right\}, \mathcal {K}} \right) {\text{ if }} \nu \tilde{n}.\mathcal {K} \vdash M \text{ and } \nu \tilde{n}.\mathcal {K} \vdash N\hfill \\
\end{align*}}
\caption{The semantics of Statverif}\label{OSStat}
\end{figure*}

\noindent\textbf{Security Property}. StatVerif is limited to the
verification of secrecy property. The secrecy property of StatVerif
is defined as follows.

\noindent\textbf{Definition 13 (StatVerif Secrecy Property)}. Let
$P$ be a closed StatVerif process, $M$ a message. $P$ preserves the secrecy of
$M$ if there exists no trace of the form:
\[
\begin{gathered}
(\emptyset,\emptyset, \{(P,0)\}, fn(P))\xrightarrow{{\alpha}}^* (\tilde{n}, \mathcal {S}, \mathcal {P}, \mathcal {K}) \text{ where } \nu \tilde{n}.\mathcal {K} \vdash M\\
\end{gathered}
\]

In the following, we encode the StatVerif processes into a subset of
SAPIC processes and obtain the CS result directly from that of
SAPIC, which has been proved in Section 3.2. With this encoding, we
can easily embed the StatVerif calculus into the CoSP framework.
Thus we do not need to build another computational execution model
for StatVerif like what we have done for SAPIC.

There are many differences between the semantics of these two
calculi. The lock construct is the place in which they
differ the most. For a StatVerif process $P:= \text{lock};P_1$, it
will lock the state and all the processes in parallel cannot access
the current state cells until an unlock in $P_1$ is achieved. For a
SAPIC process $P:= \text{lock }M;P_1$, it will only store the
$\pi$-term $M$ in a set $\Lambda$ and make sure it cannot be locked
again in another concurrent process $Q:= \text{lock }M';Q_1$ where
$M'=_E M$ until an unlock construct is achieved. Moreover, the state
cells in StatVerif calculus should be initialized before they can be
accessed. It is not required in SAPIC. Thus we should do more for a
SAPIC process to simulate the state construct in a StatVerif
process.

\begin{figure*}
\[\begin{array}{l}
 {\left\lfloor 0 \right\rfloor _0} = 0 \quad  {\left\lfloor {P|Q} \right\rfloor _0} = {\left\lfloor P \right\rfloor _0}|{\left\lfloor Q \right\rfloor _0}\quad {\left\lfloor {\nu n;P} \right\rfloor _b} = \nu n;{\left\lfloor P \right\rfloor _b}\quad {\left\lfloor {!P} \right\rfloor _0} = !{\left\lfloor P \right\rfloor _0} \\
 {\left\lfloor {{\text{in}}\left( {M,x} \right);P} \right\rfloor _b} = {\text{in}}\left( {M,x} \right);{\left\lfloor P \right\rfloor _b}\quad {\left\lfloor {{\text{out}}\left( {M,N} \right);P} \right\rfloor _b} = {\text{out}}\left( {M,N} \right);{\left\lfloor P \right\rfloor _b} \\
 {\left\lfloor {{\text{let }}x = D{\text{ in }}P{\text{ else }}Q} \right\rfloor _b} = {\text{let }}x = D{\text{ in }}{\left\lfloor P \right\rfloor _b}{\text{ else }}{\left\lfloor Q \right\rfloor _b} \quad {\left\lfloor {{\text{event }}e;P} \right\rfloor _b} = {\text{event }}e;{\left\lfloor P \right\rfloor _b} \\
 {\left\lfloor {[s \mapsto M]} \right\rfloor _0} = {\text{insert }}s,M \\
 {\left\lfloor {{\text{lock}};P} \right\rfloor _0} = {\text{lock }}l;{\left\lfloor P \right\rfloor _1}\quad {\left\lfloor {{\text{unlock}};P} \right\rfloor _1} = {\text{unlock }}l;{\left\lfloor P \right\rfloor _0} \\
 {\left\lfloor {s: = M;P} \right\rfloor _b} = \left\{ {\begin{array}{*{20}{r}}
   {{\text{lock }}l;{\text{lookup }}s{\text{ as }}{x_s}{\text{ in insert }}s,M;{\text{unlock }}l;{{\left\lfloor P \right\rfloor }_0}{\text{ for }}b = 0}\\
   {{\text{lookup }}s{\text{ as }}{x_s}{\text{ in insert }}s,M;{{\left\lfloor P \right\rfloor }_1}{\text{ for }}b = 1}  \\
    \text{where } x_s \text{ is a fresh variable}\\
\end{array}} \right. \\
 {\left\lfloor {{\text{read }}s{\text{ as }}x;P} \right\rfloor _b} = \left\{ {\begin{array}{*{20}{r}}
   {{\text{lock }}l;{\text{lookup }}s{\text{ as }}x{\text{ in unlock }}l;{{\left\lfloor P \right\rfloor }_0}{\text{ for }}b = 0}  \\
   {{\text{lookup }}s{\text{ as }}x{\text{ in }}{{\left\lfloor P \right\rfloor }_1}{\text{ for }}b = 1}  \\
\end{array}} \right. \\
 \end{array}\]
\caption{Encoding Statverif process}\label{ESP}
\end{figure*}

We first define the encoding $\lfloor P \rfloor_b$ for StatVerif
process $P$ with the boolean $b$ indicating whether $P$ has locked
the state. Note that we only need to encode the StatVerif state
constructs by using SAPIC functional state constructs. We leave the
standard constructs unchanged. For the sake of completeness, we list
them all in \textbf{Fig. \ref{ESP}}. The state cell initialization
$[s\mapsto M]$ is represented by the construct insert $s,M$. To
encode the lock operation, we set a free fresh cell name $l$. The
lock is represented by lock $l$ and turning the boolean $b$ from $0$
to $1$. The unlock construct is done in the opposite direction. To
write a new value into an unlocked state cell ($s:=M$ for $b=0$), we
need to perform 4 steps. We first lock $l$ before the operation. It
is to ensure the state is not locked in concurrent processes. We
then read the original value in $s$ to ensure $s$ has been
initialized. We complete the writing operation by the construct
insert $s,M$ and finally unlock $l$. When the state has been locked
($s:=M$ for $b=1$), we omit the contructs lock $l$ and unlock $l$
because it has been locked before and the boolean $b$ could be
turned from $1$ to $0$ only by an unlock construct. The reading
operation is similar where we bind the value to $x$ instead of a
fresh variable $x_s$.

Let $O = \left( {\tilde n,\mathcal {S}, \mathcal {P},\mathcal {K}}
\right)$ be a StatVerif semantic configuration where $\mathcal
{P}=\{(P_i, \beta_i)\}_{i=1}^k$ and $\beta_i \in \{0,1\}$ indicating
whether $P_i$ has locked the state. We define the encoding $\lfloor
O \rfloor$ as SAPIC semantic configuration.

\[\begin{array}{c}
\left\lfloor O \right\rfloor  = \left\{ {\begin{array}{*{20}{r}}
   {\left( {\tilde n,\mathcal {S},\emptyset , \{\lfloor P_i \rfloor_{\beta_i}\}_{i=1}^k , \mathcal {K},\left\{ l \right\}} \right)} {\text{ if }} \exists (P_i,\beta_i ) \in \mathcal {P} ,\beta_i  = 1, \\
   {\left( {\tilde n,\mathcal {S},\emptyset , \{\lfloor P_i \rfloor_{\beta_i}\}_{i=1}^k , \mathcal {K},\emptyset } \right)}  {\text{ if }} \forall (P_i,\beta_i ) \in \mathcal {P} ,\beta_i  = 0.\\
\end{array}} \right.
 \end{array}\]

Before we prove Lemma 6 that our encoding is able to capture
secrecy of StatVerif process, we provide Lemma 4 and Lemma 5 to
explain that the encoding SAPIC process can simulate the encoded
StatVerif process. Then by Theorem 2 we obtain the CS result of
StatVerif. The proofs are in Appendix B.

\noindent\textbf{Lemma 4}. Let $O_1$ be a StatVerif semantic
configuration. If $O_1 \xrightarrow{\alpha} O_2$, then $\lfloor
O_1\rfloor \xrightarrow{\alpha}^* \lfloor O_2\rfloor$.

\noindent\textbf{Lemma 5}. Let $O_1$ be a StatVerif semantic
configuration. If $\lfloor O_1\rfloor \xrightarrow{\alpha} O'$, then
there exists a StatVerif semantic configuration $O_2$, such that
$O_1 \xrightarrow{\alpha}^*  O_2$ and that $O' = \lfloor O_2\rfloor$
or $O' \xrightarrow{}^* \lfloor O_2\rfloor$.

\noindent\textbf{Lemma 6}. Let $P_0$ be a closed StatVerif process.
Let $M$ be a message. Set $P':=\text{in}(attch, x);\text{let
}y=equal(x,M)$ in event $NotSecret $, where $x,y$ are two fresh
variables that are not used in $P_0$, $attch \in \textbf{N}_E$ is a
free channel name which is known by the adversary. We set $\wp
:=\{e|NotSecret$ is not in $e\}$. $Q_0:=\lfloor P'|P_0\rfloor_0$ is
a closed SAPIC process and $\wp $ is a SAPIC trace property. Then we
have that $P_0$ symbolically preserves the secrecy of $M$ (in the
sense of Definition 13) \emph{iff} $Q_0$ symbolically satisfies $\wp
$ (in the sense of Definition 7).

\noindent\textbf{Theorem 2 (CS in StatVerif)}. Assume that the
computational implementation of the applied $\pi$ calculus is a
computationally sound implementation (Definition 3) of the symbolic
model of the applied $\pi$ calculus (Definition 4) for a class
\textbf{P} of protocols. For a closed StatVerif process $P_0$, we
denote by $Q_0$ and $\wp$ the same meanings in Lemma 6. Thus if the
StatVerif process $P_0$ symbolically preserves the secrecy of a
message $M$ (Definition 13) and $\Pi_{Q_0}^S\in\textbf{P}$, then
$Q_0$ computationally satisfies $\wp$.

\section{Case Study: CS Results of Public-Key Encryption and Signatures}

In section 3 and 4, we have embedded the stateful applied $\pi$
calculus used in SAPIC and StatVerif into the CoSP framework. CoSP
allows for casting CS proofs in a conceptually modular and generic
way: proving $x$ cryptographic primitives sound for $y$ calculi only
requires $x+y$ proofs (instead of $x\cdot y$ proofs without this
framework). In particular with our results, all CS proofs that have
been conducted in CoSP are valid for the stateful applied $\pi$
calculus, and hence accessible to SAPIC and StatVerif.

We exemplify our CS results for stateful applied $\pi$ calculus by
providing the symbolic model that is accessible to the two
verification tools, SAPIC and StatVerif. We use the CS proofs in
\cite{Computational12} with a few changes fitting for the
verification mechanism in these tools. The symbolic model allows for
expressing public-key encryption and signatures.

Let $\textbf{C}:=\{ enc/3,$ $ ek/1,$ $ dk/1,$ $ sig/3,$ $ vk/1,$ $
sk/1,$ $ pair/2,$ $ string_0/1,$ $ string_1/1,$ $ empty/0,$ $
garbageSig/2,$ $ garbage/1,$ $ garbageEnc/2\}$ be the set of
constructors. We require that $ \textbf{N} = \textbf{N}_\textbf{E}
\uplus \textbf{N}_\textbf{P}$ for countable infinite sets
$\textbf{N}_\textbf{P}$ of protocol nonces and
$\textbf{N}_\textbf{E}$ of attacker nonces. Message type \textbf{T}
is the set of all terms $T$ matching the following grammar, where
the nonterminal $N$ stands for nonces.
\begin{align*}
T::= & enc(ek(N), T, N) | ek(N) | dk(N) | sig(sk(N), T, N) |vk(N)|sk(N) |\\
     & pair(T, T) | S | N | garbage(N) | garbageEnc(T, N) |garbageSig(T, N)\\
S::= &empty | string_0(S) | string_1(S)\\
\end{align*}
\noindent Let $\textbf{D}:=\{dec/2,$ $isenc/1,$ $isek/1,$ $isdk/1,$
$ekof/1,$ $ekofdk/1,$ $verify/2,$ $issig/1,$ $isvk/1,$ $issk/1,$
$vkof/2,$ $vkofsk/1,$ $fst/1,$ $snd/1,$ $unstring_0/1,$ $equal/2\}$
be the set of destructors. The full description of all destructor
rules is given in \cite{Computational12}. Let $\vdash$ be defined as
in Definition 4. Let
$\textbf{M}=(\textbf{C},\textbf{N},\textbf{T},\textbf{D},\vdash)$ be
the symbolic model.

In StatVerif, the symbolic model \textbf{M} can be directly achieved
since the term algebra is inherited from ProVerif, whose CS property
has been proved in \cite{Computational12}. In SAPIC, we formalize
the symbolic model by a signature $\Sigma := \textbf{C} \cup
\textbf{D}$ with the equational theories expressing the destructor
rules. Note that 3 destructor rules are filtered out including: i)
$ekofdk(dk(t)) = ek(t)$; ii) $vkof(sig(sk(t_1), t_2, t_3)) =
vk(t_1)$; iii) $vkofsk(sk(t)) = vk(t)$, since they are not
subterm-convergent, which is required by SAPIC (by verification
mechanism of tamarin-prover). Note that these rules are all used to
derive the public key. We require that for all the signatures and
private keys in communication, they should be accompanied by their
public keys. In this way, both the adversary and the protocol will
not use these rules. To show the usefulness of our symbolic model in
this section, we have verified the left-or-right protocol presented
in \cite{Statverif} by using SAPIC and StatVerif. In Appendix C and D, we provide
the scripts for the protocol.

To establish CS results, we require the protocols to fulfill several
natural conditions with respect to their use of randomness.
Protocols that satisfy these protocol conditions are called
\emph{randomness-safe}. Additionally, the cryptographic
implementations needs to fulfill certain conditions, e.g., that the
encryption scheme is PROG-KDM secure, and the signature scheme is
SUF-CMA. Both the protocol conditions and the implementation
conditions could be found in \cite{Computational12}. Then we
conclude CS for protocols in the stateful applied $\pi$ calculus
that use public-key encryption and signatures.

\noindent\textbf{Theorem 3 (CS for Enc. and Signatures in SAPIC
and StatVerif)}. Let \textbf{M} be as defined in this
section and $A$ of \textbf{M} be an implementation that satisfies
the conditions from above. If a randomness-safe closed SAPIC or StatVerif
process $P_0$ symbolically satisfies a trace property $\wp$,
then $P_0$ computationally satisfies $\wp$\footnote{For a closed StatVerif process $P_0$,
we denote by $Q_0$ and $\wp$ the same meanings in Lemma 6. We say $P_0$ computationally
satisfies $\wp$ iff $Q_0$ computationally satisfies $\wp$.}.

\section{Conclusion}

In this paper, we present two CS results respectively for the
stateful applied $\pi$ calculus used in SAPIC tool and StatVerif
tool. We show that the CS results of applied $\pi$ calculus implies
the CS results of SAPIC calculus and of StatVerif calculus. Thus for
any computationally sound implementation of applied $\pi$ calculus,
if the security property of a closed stateful process is verified by
SAPIC tool or StatVerif tool, it is also computationally satisfied.
The work is conducted within the CoSP framework. We give the
embedding from the SAPIC calculus to CoSP protocols. Furthermore, we
provide an encoding of the StatVerif processes into a subset of
SAPIC processes, which shows the differences between the semantics
of these two calculi. As a case study, we provide the CS result for
the input languages of StatVerif and SAPIC with public-key
encryption and signatures.

%\section{Acknowledgments}
%
%The research presented in this paper is supported by the National
%Basic Research Program of China (No. 2013CB338003) and National
%Natural Science Foundation of China (No. 91118006, No.61202414).

\newpage
\makeatletter
\renewcommand\@biblabel[1]{{[#1]\hfill}}
\makeatother

%\bibliographystyle{ieeetr}
%\bibliography{ref}

\newpage

\section*{Appendix A: Proof of Theorem 1}
\noindent\textbf{Lemma 1}. $SExec_{P_0}^S$ satisfies a trace
property $\wp $ \emph{iff} $\Pi_{P_0}^S$ satisfies $events^{-1}(\wp
)$. Moreover, $P_0$ computationally satisfies $\wp $ \emph{iff}
$(\Pi_{P_0}^S, A)$ computationally satisfies $events^{-1}(\wp )$.
Both are in the sense of Definition 2.

\begin{proof}
Since $events^{-1}(\wp )$ is the set of nodes sequences whose raised
events sequence are in the trace property set $\wp $. Thus the
symbolic case is immediate from the construction of $\Pi_{P_0}^S$.
For the computational case, note that the computational
implementation of $P_0$ is defined like the symbolic one, except
that it uses the implementations of the CoSP-constructors (includes
the nonces) and CoSP-destructors (includes destructor $equal'$)
rather than operate abstractly on CoSP-terms. Thus it is true for
the computational case.
\end{proof}

\noindent\textbf{Lemma 2}. The CoSP protocol $\Pi_{P_0}^S$ is
efficient.

\begin{proof}
By construction, given all the node
identifiers and the edge labels on the path to a node $N$, there
should be a deterministic polynomial-time algorithm that can compute
the label of $N$ (the current state of the CoSP protocol). According
to the construct of $SExec_{P_0}^S$, it only needs to prove that the
computation steps in each loop is bounded by a polynomial in the
loop number. We only consider the state constructs and others are
with constant numbers. For the functional state constructs, the
number of computation steps in set-membership decision $f_{mem}$ is
bounded by the cardinal number of set, which is less than the
reduction steps of main process. For the multiset state construct,
as we have stated in Section 3.2 that the time complexity for the
pattern matching algorithm $f_{match}$ is polynomial in the
reduction steps. Thus the computation steps of the algorithm would
not grow exponentially.

It is left to show that the length of the node identifier is bounded
by a polynomial in the length of the path leading to that node. This
is equivalent to showing that the state tuple $(\mathcal {P}, \mu,
\eta, S,S^{MS}, \Lambda)$ of $SExec_{P_0}^S$ is of polynomial-length
(when not counting the length of the representations of the
CoSP-terms). For $\mu$, $\eta$, $S$, and $\Lambda$, this is
immediately satisfied since they grow by at most one entry in each
activation of $SExec_{P_0}^S$. For $S^{MS}$, we have stated in
Section 3.2 that its size is polynomial in the number of reduction
steps since we treat the CoSP-terms as black-boxes. At last, we
should show that the size of processes $P$ in $\mathcal {P}$ is
polynomially bounded. Note the fact that in each activation of
$SExec_{P_0}^S$, processes $P$ either gets smaller, or we have
$\mathcal {P}=E[!P_1]$ and processes $P$ in $\mathcal {P}$ grow by
the size of $P_1$, which is bounded by the size of $P_0$. Thus the
size of processes $P$ in $\mathcal {P}$ is linear in the number of
activation of $SExec_{P_0}^S$.
\end{proof}

\noindent\textbf{Lemma 3 (Safe Approximation for SAPIC)}. If a SAPIC
closed process $P_0$ symbolically satisfies a SAPIC trace property
$\wp $ in the sense of Definition 7, then $SExec_{P_0}^S$ satisfies
$\wp $ in the sense of Definition 12.

\begin{proof}.
To show this lemma, it is sufficient to show that if $SExec_{P_0}^S$
raises events $e_1,...,e_n$, then $\underline{e}$ is a SAPIC event
trace of $P_0$. Hence, for the following we fix an execution of
$SExec_{P_0}^S$ in interaction with a Dolev-Yao adversary $\mathcal
{A}$ in which $SExec_{P_0}^S$ raises the events $e_1,...,e_n$. We
then prove the lemma by showing that there exists a finite sequence
$[\mathcal {K}_1, ..., \mathcal {K}_n]$ of sets of $\pi$-terms such
that
$(\emptyset,\emptyset,\emptyset,\{P_0\},fn(P_0),\emptyset)\xrightarrow{}^*\xrightarrow{e_1}(\tilde
n_1,\mathcal {S}_1, \mathcal{S}^{MS}_1,\mathcal{P}_1,\mathcal
{K}_1,\mathcal{L}_1)\xrightarrow{}^*\xrightarrow{e_2}\cdots\xrightarrow{}^*\xrightarrow{e_n}\\(\tilde
n_m,\mathcal {S}_m, \mathcal{S}^{MS}_m,\mathcal{P}_m,\mathcal
{K}_m,\mathcal{L}_m).$

For a given iteration of the main loop of $SExec_{P_0}^S$, let
$(\mathcal {P}, \eta, \mu, S, S^{MS}, \Lambda)$ denote the
corresponding state of $SExec_{P_0}^S$ at the beginning of that
iteration. Let $E$ denote the evaluation context chosen in that
iteration. Let $\underline{n}$ be the domain of $\mu$ without the
names $r_1,...,r_n$ sent in the very beginning of the execution of
$SExec_{P_0}^S$. $(\mathcal {P}', \eta', \mu', S',S^{MS'},
\Lambda')$ and $\underline{n}'$ are the corresponding values after
that iteration. Let $fromadv$ be the list of terms received from the
adversary in that iteration, and let $toadv$ be the list of terms
sent to the adversary. By $(\mathcal {P}_0, \eta_0, \mu_0, S_0,
S^{MS}_0, \Lambda_0)$ we denote the corresponding values before the
first iteration but after the sending of the message $(r_1,
...,r_n)$, and by $(\mathcal {P}_*, \eta_*, \mu_*, S_*, S^{MS}_*,
\Lambda_*)$ and $\underline{n}_*$ the values after the last
iteration. We call a variable or name \emph{used} if it occurs in
the domain of $\eta_*$ or $\mu_*$, respectively. Note that
$\mu_0=(a_1\mapsto r_1,...,a_n\mapsto r_n)$ where $\underline{a}$
are the free names in $P_0$, but $\underline{n}_0=\emptyset$. Note
that $\mathcal {P}$ will never contain unused free variables or
names.

Let $K$ denote the list of all CoSP-terms output by $SExec_{P_0}^S$
up to the current iteration. We encode $K=(t_1, ..., t_m)$ as a
substitution $\varphi$ mapping $x_i\mapsto t_i$ where $x_i$ are
arbitrary unused variables. We denote by $K',\varphi'$,
$K_0,\varphi_0$ and $K_*,\varphi_*$ the values of $K,\varphi$ after
the current iteration, before the iteration (but after sending
$(r_1, ...,r_n)$), and after the last iteration, respectively. Note
that $K_0=(r_1, ...,r_n)$.

Let $\gamma$ be an injective partial function that maps every $N\in
\textbf{N}_E$ to an unused $\pi$-name, and every $N\in \text{range
}\mu_*$ to $\mu_*^{-1}(N)$. (This is possible because $\text{range
}\mu_*\subseteq\textbf{N}_P$ and $\mu_*$ is injective.) We
additionally require that all unused $\pi$-names are in $\text{range
}\gamma$. (This is possible since both $\textbf{N}_E$ and the set of
unused $\pi$-names are countably infinite.)

The following claims proposed in \cite{CoSP} can still stick. Note
that for any $\pi$-destructor $d$ and any $\pi$-terms
$\underline{M}$ with $fv(\underline{M})\in dom(\eta)$ and
$fn(\underline{M})\in dom(\mu)$, we have that $M\eta\mu$ are
CoSP-terms and $d'(M\eta\mu)\gamma = d(M\eta\mu\gamma)$ (where $d'$
is as in Section 2.2). Hence for a destructor term $D$ with
$fv(D)\subseteq dom(\eta)$ and $fn(D)\subseteq dom(\mu)$, we have
$eval^{CoSP}(D\eta\mu)\gamma=eval^\pi(D\eta\mu\gamma)$. Since
$a\mu\gamma=a$ for all names $a\in dom(\mu)$, $D\eta\mu\gamma
= D\eta\gamma$. Since $eval^{CoSP}(D\eta\mu)$ does not contain
variables, $eval^{CoSP}(D\eta\mu)=eval^{CoSP}(D\eta\mu)\eta$. Thus
for $D$ with $fv(D)\subseteq dom(\eta)$ and
$fn(D)\subseteq dom(\mu)$, we have
\begin{align}
eval^{CoSP}(D\eta\mu)\eta\gamma=eval^{\pi}(D\eta\gamma).
\end{align}
\noindent where the left hand side is defined iff the right hand
side is defined.

Similarly to (1), if $fv(D)\subseteq dom(\varphi)$ and
$fn(D)\subseteq dom(\gamma^{-1})$, we have that
$eval^{CoSP}(D\varphi\gamma^{-1})\gamma=eval^{\pi}(D\varphi\gamma)$.
For a CoSP-term $t$ with $K\vdash t$, from the definition of
$\vdash$ it follows that $t = eval^{CoSP}(D_t\varphi\gamma^{-1})$
for some destructor $\pi$-term $D_t$ containing only unused names
and variables in $dom(\varphi)$ (note that every $N \in
\textbf{N}_E$ can be expressed as $a\gamma^{-1}$ for some unused
$a$). Since all unused names are in $dom(\gamma^{-1})$, we
have
\begin{align}
t\gamma=eval^{CoSP}(D_t\varphi\gamma^{-1})\gamma=eval^{\pi}(D_t\varphi\gamma).
\end{align}

Given two CoSP-terms $t, u$ such that $equal'(t,u)\neq\bot$ and $t$
and $u$ only contain nonces $N\in \textbf{N}_E\cup\text{range
}\mu_*$, by definition of $equal'$ and using that
$\gamma$ is injective and defined on $ \textbf{N}_E\cup\text{range
}\mu_*$, we have $equal'(t,u)=equal(t\gamma, u\gamma)\gamma^{-1}$
and hence $equal(t\gamma, u\gamma)\neq \bot$. Hence, for $t,u$ only
containing nonces $N\in \textbf{N}_E\cup\text{range }\mu_*$, we have
that
\begin{align}
equal'(t,u)\neq \bot \Leftrightarrow t\gamma=_E u\gamma
\end{align}

%For $D$ with $fv(D)\subseteq\text{dom }\eta$ and
%$fn(D)\subseteq\text{dom }\mu$, we have
%\[
%eval^{CoSP}(D\eta\mu)\eta\gamma=eval^{\pi}(D\eta\gamma).
%\]
%\noindent where the left hand side is defined \emph{iff} the right
%hand side is.
%
%For a CoSP-term $t$ with $K\vdash t$, there exists a destructor
%$\pi$-term $D_t$ containing only unused names and variables in
%$\text{dom }\varphi$ such that
%\[
%t\gamma=eval^{CoSP}(D_t\varphi\gamma^{-1})\gamma=eval^{\pi}(D_t\varphi\gamma).
%\]
%
%For $t,u$ only containing nonces $N\in \textbf{N}_E\cup\text{range
%}\mu_*$, we have that
%\[
%equal'(t,u)\neq \bot \Leftrightarrow t\gamma=_E u\gamma
%\]

\noindent\textbf{Claim:} The main loop in $SExec_{P_0}^S$ satisfies that
$(\underline{n}, S\eta\gamma, S^{MS}\eta\gamma,\mathcal
{P}\eta\gamma, K\gamma, \Lambda\eta\gamma)\Rightarrow
(\underline{n}', S'\eta'\gamma, S^{MS'}\eta'\gamma,\mathcal
{P}'\eta'\gamma, K'\gamma, \Lambda'\eta'\gamma)$. Here $\Rightarrow$
denotes $\xrightarrow{e}$ if an event $e$ is raised in the current
iteration, and $\xrightarrow{}^*$ otherwise.
$S\eta\gamma:=\{(r\eta\gamma, r'\eta\gamma)| \forall (r, r') \in
S\}$. $S^{MS}\eta\gamma$ is similar by applying the mapping
$\eta\gamma$ to all the arguments of facts in $S^{MS}$.

Assuming that we have shown this claim, it follows that
$(\underline{n}_0, S_0\eta_0\gamma, S^{MS}_0\eta\gamma,\\ \mathcal
{P}_0\eta_0\gamma, K_0\gamma,
\Lambda_0\eta_0\gamma)\xrightarrow{}^*\xrightarrow{e_1}\xrightarrow{}^*
\xrightarrow{e_2}\xrightarrow{}^*\cdots
\xrightarrow{}^*\xrightarrow{e_n} $. Since
$\eta_0=S_0=S^{MS}=\underline{n}_0= \Lambda_0=\emptyset$,
$K_0\gamma=\underline{a}=fn(P_0)$ and since $\mathcal {P}_0=\{P_0\}$
does not contain nonces $N\in \textbf{N}$, we have $\mathcal
{P}_0\eta_0\gamma = \mathcal {P}_0$. Then we have that $(\emptyset,
\emptyset,\emptyset, \{P_0\},fn(P_0),\emptyset)
\xrightarrow{}^*\xrightarrow{e_1}\xrightarrow{}^*\xrightarrow{e_2}\xrightarrow{}^*\cdots\xrightarrow{}^*\xrightarrow{e_n}$.
This implies that $\underline{e}$ is a SAPIC event trace of $P_0$.
It proves this lemma.

It is left to prove the claim. We distinguish the following cases:

i) In the following cases, the adversary chooses to proceed the
standard $\pi$-process except for the input and output constructs
where $\mathcal {P}=E[\nu a;P_1]$, or $\mathcal
{P}=E[\text{out}(M_1,N);P_1][\text{in}(M_2,x);P_2]$, or $\mathcal
{P}=E[\text{let}{\kern 3pt}x=D{\kern 3pt}\text{in}{\kern
3pt}P_1{\kern 3pt} \text{else}{\kern 3pt}P_2]$, or $\mathcal
{P}=E[\text{event}{\kern 3pt}e;P_1]$, or $\mathcal {P}=E[!P_1]$. In
these cases, we have $S'=S$, $S^{MS'}=S^{MS}$, $K'=K$,
$\Lambda'=\Lambda$. For all $x\in dom(\eta')\backslash dom(\eta)$,
we have $x \notin fv(S)\cup fv(\Lambda)\cup fv(S^{MS})$. Thus
$S\eta\gamma=S'\eta'\gamma$, $\Lambda\eta\gamma=\Lambda'\eta'\gamma$
and $S^{MS}\eta\gamma=S^{MS'}\eta'\gamma$. According to the proof of
Lemma 4 in \cite{CoSP}, we have that $(\underline{n}, S\eta\gamma,
S^{MS}\eta\gamma,\mathcal {P}\eta\gamma, K\gamma,
\Lambda\eta\gamma)\Rightarrow (\underline{n}', S'\eta'\gamma,
S^{MS'}\eta\gamma,\mathcal {P}'\eta'\gamma, K'\gamma,
\Lambda'\eta'\gamma)$.

ii) $\mathcal {P}=E[\text{in}(M,x);P_1]$ and $fromadv=(c,m)$ and
$eval^{CoSP}M\eta\mu \cong c$: Then $\mathcal {P}'=E[P_1]$, $K'=K$,
$S'=S$, $S^{MS'}=S^{MS}$, $\Lambda'=\Lambda$, $\mu'=\mu$, and
$\eta'=\eta\cup\{x:=m\}$. Furthermore, since $SExec_{P_0}^S$
interacts with a Dolev-Yao adversary, $K\vdash c,m$. By (2), we have
$K\gamma \vdash c\gamma, m\gamma$. Since a Dolev-Yao adversary will
never derive protocol nonces that have never been sent, we have that
only nonces $N\in \textbf{N}_E \cup \text{range }\mu$ occur in $c$
and in $M\eta\mu$. Hence with (3), from $M\eta\mu =
eval^{CoSP}M\eta\mu \cong c$ it follows that
$M\eta\gamma=M\eta\mu\gamma =_E c\gamma$. Thus we have
\begin{align*}
  &\left( {\underline{n},S\eta \gamma , S^{MS}\eta\gamma, \mathcal {P} \eta \gamma, K\gamma, \Lambda\eta\gamma } \right) \\
= &\left( {\underline{n},S\eta \gamma , S^{MS}\eta\gamma,\left( {E\eta \gamma } \right)\left[ \text{in}(M\eta\gamma,x);P_1\eta\gamma \right], K\gamma, \Lambda\eta\gamma} \right) \\
\to &\left( {\underline{n},S\eta \gamma,S^{MS}\eta\gamma,\left( {E\eta \gamma } \right)\left[ P_1\eta\gamma\{m\gamma/x\} \right], K\gamma, \Lambda\eta\gamma} \right) \\
= &\left( {\underline{n}',S'\eta' \gamma , S^{MS'}\eta'\gamma, \mathcal {P}' \eta' \gamma, K'\gamma, \Lambda'\eta'\gamma } \right) \\
\end{align*}
\noindent Since we maintain the invariant that all bound variables
in $\mathcal {P}$ are distinct from all other variables in $\mathcal
{P}$, or $S$, or $\lambda$, or $dom(\eta)$, we have $x\notin fv(E)$,
$x\notin S$, $x\notin \lambda$, $x\notin S^{MS}$, and $x\notin
dom(\eta)$. Hence $E\eta \gamma  = E\eta '\gamma$, ${P_1}\eta \gamma
\{ m\gamma /x \} = {P_1}\eta \{ {m/x} \}\gamma  = {P_1}\eta
'\gamma$, $\Lambda \eta \gamma  = \Lambda \eta '\gamma $. Thus the last equation
is true.

iii) $\mathcal {P}=E[\text{out}(M,N);P_1]$ with $t_M':=fromadv\cong
t_M$ and $toadv=t_N$ where $t_M\cong eval^{CoSP}(M\eta\mu)$ and
$t_N:=eval^{CoSP}(N\eta\mu)$. Then $K'=K\cup\{t_N\}$, $\mathcal
{P}'=E[P_1]$, $\eta'=\eta$, $\mu'=\mu$, $S'=S$, $S^{MS'}=S^{MS}$,
and $\Lambda'=\Lambda$. Since $t_M'$ was sent by the adversary,
$K\vdash t_M'$. According to the Dolev-Yao property, the adversary
will never derive protocol nonces that have never been sent, we have
that only nonces $N\in \textbf{N}_E\cup \text{range }\mu$ occur in
$t_M'$ and $M\eta\mu$. Hence with (3), from $t_M'\cong t_M$ it
follows that $M\eta\gamma=M\eta\mu\gamma=t_M\gamma=_E t_M'\gamma$,
and that $N\eta\gamma=t_N\gamma$. Thus $K\gamma \vdash M\eta\gamma$,
and $K'\gamma=K\gamma\cup\{t_N\gamma\}=K\gamma \cup
\{N\eta\gamma\}$. We have that
\begin{align*}
  &\left( {\underline{n},S\eta \gamma , S^{MS}\eta\gamma, \mathcal {P} \eta \gamma, K\gamma, \Lambda\eta\gamma } \right) \\
= &\left( {\underline{n},S\eta \gamma , S^{MS}\eta\gamma,\left( {E\eta \gamma } \right)\left[ \text{out}(M\eta\gamma, N\eta\gamma);P_1\eta\gamma \right], K\gamma, \Lambda\eta\gamma} \right) \\
\to &\left( {\underline{n},S\eta \gamma,S^{MS}\eta\gamma,\left( {E\eta \gamma } \right)\left[ P_1\eta\gamma \right], K\gamma \cup\{N\eta\gamma\}, \Lambda\eta\gamma} \right) \\
= &\left( {\underline{n}',S'\eta' \gamma , S^{MS'}\eta'\gamma, \mathcal {P}' \eta' \gamma, K'\gamma, \Lambda'\eta'\gamma } \right) \\
\end{align*}

iv) $\mathcal {P} = E[\text{insert }M,N;P_1]$ with
$t_M=eval^{CoSP}(M\eta\mu)$, $t_N=eval^{CoSP}(N\eta\mu)$, and
$\exists (r, r')\in S$ such that $r\cong t_M$. Then $K' = K,\mathcal
{P}'=E[P_1],\eta' = \eta ,\mu' = \mu, S' = S \backslash \{(r, r')\}
\cup\{(t_M, t_N)\}, S^{MS'} = S^{MS},\Lambda' = \Lambda$. By (1), $
t_M\eta\gamma = eval^{\pi}(M\eta\gamma) = M\eta\gamma$,
$t_N\eta\gamma = N\eta\gamma$. $r\cong t_M$ implies $r\eta \cong
t_M\eta$. Since a Dolev-Yao adversary will never derive protocol
nonces that have never been sent, we have that only nonces $N\in
\textbf{N}_E \cup \text{range }\mu$ occur in $r\eta$ and $t_M\eta$.
By (3), $M\eta\gamma = t_M\eta\gamma= r\eta\gamma \in
dom(S)\eta\gamma$. Thus we have
\begin{align*}
  &\left( {\underline{n},S\eta \gamma , S^{MS}\eta\gamma, \mathcal {P} \eta \gamma, K\gamma, \Lambda\eta\gamma } \right) \\
= &\left( {\underline{n},S\eta \gamma , S^{MS}\eta\gamma,\left( {E\eta \gamma } \right)\left[ \text{insert }M\eta\gamma, N\eta\gamma;P_1\eta\gamma \right], K\gamma, \Lambda\eta\gamma} \right) \\
\to &( \underline{n},S\eta \gamma\backslash\{(r\eta\gamma,r'\eta\gamma) \}\cup\{M\eta\gamma \mapsto N\eta\gamma\},S^{MS}\eta\gamma,\left( {E\eta \gamma } \right)\left[ P_1\eta\gamma \right], K\gamma , \Lambda\eta\gamma ) \\
= &( \underline{n}',\left(S\backslash\{(r,r')\}\cup\{(t_M, t_N)\}\right) \eta\gamma , S^{MS'}\eta'\gamma,) \mathcal {P}' \eta' \gamma, K'\gamma, \Lambda'\eta'\gamma  ) \\
= &\left( {\underline{n}',S'\eta' \gamma , S^{MS'}\eta'\gamma, \mathcal {P}' \eta' \gamma, K'\gamma, \Lambda'\eta'\gamma } \right) \\
\end{align*}

v) $\mathcal {P} = E[\text{insert }M,N;P_1]$ with
$t_M=eval^{CoSP}(M\eta\mu)$, $t_N=eval^{CoSP}(N\eta\mu)$, and
$\forall (r,r')\in S$, $r\ncong t_M$. Then $K' = K,\mathcal
{P}'=E[P_1],\eta' = \eta ,\mu' = \mu, S' = S \cup\{(t_M, t_N)\},
S^{MS'} = S^{MS},\Lambda' = \Lambda$. By (1), $ t_M\eta\gamma =
eval^{\pi}(M\eta\gamma) = M\eta\gamma$, $t_N\eta\gamma =
N\eta\gamma$. $r\ncong t_M$ implies $r\eta \ncong t_M\eta$. Since a
Dolev-Yao adversary will never derive protocol nonces that have
never been sent, we have that only nonces $N\in \textbf{N}_E \cup
\text{range }\mu$ occur in $r\eta$ and $t_M\eta$. By (3),
$M\eta\gamma = t_M\eta\gamma \neq r\eta\gamma$ for all $ r\eta\gamma
\in dom(S)\eta\gamma$. Thus we have
\begin{align*}
  &\left( {\underline{n},S\eta \gamma , S^{MS}\eta\gamma, \mathcal {P} \eta \gamma, K\gamma, \Lambda\eta\gamma } \right) \\
= &\left( {\underline{n},S\eta \gamma , S^{MS}\eta\gamma,\left( {E\eta \gamma } \right)\left[ \text{insert }M\eta\gamma, N\eta\gamma;P_1\eta\gamma \right], K\gamma, \Lambda\eta\gamma} \right) \\
\to &\left( {\underline{n},S\eta \gamma\cup\{(M\eta\gamma, N\eta\gamma)\},S^{MS}\eta\gamma,\left( {E\eta \gamma } \right)\left[ P_1\eta\gamma \right], K\gamma , \Lambda\eta\gamma} \right) \\
= &\left( {\underline{n}',\left(S\cup\{(t_M, t_N)\}\right) \eta\gamma , S^{MS'}\eta'\gamma, \mathcal {P}' \eta' \gamma, K'\gamma, \Lambda'\eta'\gamma } \right) \\
= &\left( {\underline{n}',S'\eta' \gamma , S^{MS'}\eta'\gamma, \mathcal {P}' \eta' \gamma, K'\gamma, \Lambda'\eta'\gamma } \right) \\
\end{align*}

vi) $\mathcal {P} = E[\text{delete }M;P_1]$ with
$t_M=eval^{CoSP}(M\eta\mu)$, and $\exists (r,r')\in S$ such that
$r\cong t_M$. Then $K' = K,\mathcal {P}'=E[P_1],\eta' = \eta ,\mu' =
\mu, S' = S \backslash \{(r, r')\},S^{MS'} = S^{MS}, \Lambda' =
\Lambda$. By (1), $ t_M\eta\gamma = eval^{\pi}(M\eta\gamma) =
M\eta\gamma$. $r\cong t_M$ implies $r\eta \cong t_M\eta$. Since a
Dolev-Yao adversary will never derive protocol nonces that have
never been sent, we have that only nonces $N\in \textbf{N}_E \cup
\text{range }\mu$ occur in $r\eta$ and $t_M\eta$. By (3),
$M\eta\gamma = t_M\eta\gamma= r\eta\gamma \in dom(S)\eta\gamma$.
Thus we have
\begin{align*}
  &\left( {\underline{n},S\eta \gamma , S^{MS}\eta\gamma, \mathcal {P} \eta \gamma, K\gamma, \Lambda\eta\gamma } \right) \\
= &\left( {\underline{n},S\eta \gamma , S^{MS}\eta\gamma,\left( {E\eta \gamma } \right)\left[ \text{delete }M\eta\gamma;P_1\eta\gamma \right], K\gamma, \Lambda\eta\gamma} \right) \\
\to &\left( {\underline{n},S\eta \gamma\backslash\{(r\eta\gamma, r'\eta\gamma) \},S^{MS}\eta\gamma,\left( {E\eta \gamma } \right)\left[ P_1\eta\gamma \right], K\gamma , \Lambda\eta\gamma} \right) \\
= &\left( {\underline{n}',\left(S\backslash\{(r,r')\}\right) \eta\gamma , S^{MS'}\eta'\gamma, \mathcal {P}' \eta' \gamma, K'\gamma, \Lambda'\eta'\gamma } \right) \\
= &\left( {\underline{n}',S'\eta' \gamma , S^{MS'}\eta'\gamma, \mathcal {P}' \eta' \gamma, K'\gamma, \Lambda'\eta'\gamma } \right) \\
\end{align*}

vii) $\mathcal {P} = E[\text{delete }M;P_1]$ with
$t_M=eval^{CoSP}(M\eta\mu)$, and $\forall (r,r')\in S$, $r\ncong
t_M$. Then $K' = K,\mathcal {P}'=E[P_1],\eta' = \eta ,\mu' = \mu, S'
= S,S^{MS'} = S^{MS}, \Lambda' = \Lambda$. By (1), $ t_M\eta\gamma =
eval^{\pi}(M\eta\gamma) = M\eta\gamma$. $r\ncong t_M$ implies $r\eta
\ncong t_M\eta$. Since a Dolev-Yao adversary will never derive
protocol nonces that have never been sent, we have that only nonces
$N\in \textbf{N}_E \cup \text{range }\mu$ occur in $r\eta$ and
$t_M\eta$. By (3), $M\eta\gamma = t_M\eta\gamma \neq r\eta\gamma$
for all $r\eta\gamma \in dom(S)\eta\gamma$. Thus we have
\begin{align*}
  &\left( {\underline{n},S\eta \gamma , S^{MS}\eta\gamma, \mathcal {P} \eta \gamma, K\gamma, \Lambda\eta\gamma } \right) \\
= &\left( {\underline{n},S\eta \gamma , S^{MS}\eta\gamma,\left( {E\eta \gamma } \right)\left[ \text{delete }M\eta\gamma;P_1\eta\gamma \right], K\gamma, \Lambda\eta\gamma} \right) \\
\to &\left( {\underline{n},S\eta \gamma,S^{MS}\eta\gamma,\left( {E\eta \gamma } \right)\left[ P_1\eta\gamma \right], K\gamma , \Lambda\eta\gamma} \right) \\
= &\left( {\underline{n}',S'\eta' \gamma , S^{MS'}\eta'\gamma, \mathcal {P}' \eta' \gamma, K'\gamma, \Lambda'\eta'\gamma } \right) \\
\end{align*}

viii) $\mathcal {P} = E[\text{lookup }M\text{ as }x\text{ in
}P_1\text{ else }P_2]$ with $t_M=eval^{CoSP}(M\eta\mu)$, and
$\exists (r,r')\in S$ such that $r\cong t_M$. Then $K' = K,\mathcal
{P}'=E[P_1],\eta' = \eta \cup \{x:=r'\} ,\mu' = \mu, S' = S, S^{MS'}
= S^{MS},\Lambda' = \Lambda$. By (1), $ t_M\eta\gamma =
eval^{\pi}(M\eta\gamma) = M\eta\gamma$. $r\cong t_M$ implies $r\eta
\cong t_M\eta$. Since a Dolev-Yao adversary will never derive
protocol nonces that have never been sent, we have that only nonces
$N\in \textbf{N}_E \cup \text{range }\mu$ occur in $r\eta$ and
$t_M\eta$. By (3), $M\eta\gamma = t_M\eta\gamma= r\eta\gamma \in
dom(S)\eta\gamma$. Thus we have
\begin{align*}
  &\left( {\underline{n},S\eta \gamma , S^{MS}\eta\gamma, \mathcal {P} \eta \gamma, K\gamma, \Lambda\eta\gamma } \right) \\
= &( \underline{n},S\eta \gamma , S^{MS}\eta\gamma,\left( {E\eta \gamma } \right)[ \text{lookup }M\eta\gamma\text{ as }x\text{ in }P_1\eta\gamma \text{ else }P_2\eta\gamma ], K\gamma, \Lambda\eta\gamma) \\
\to &\left( {\underline{n},S\eta \gamma ,S^{MS}\eta\gamma,\left( {E\eta \gamma } \right)\left[ P_1\eta\gamma\left\{ r'\eta\gamma/x \right\} \right], K\gamma , \Lambda\eta\gamma} \right) \\
= &\left( {\underline{n}',S'\eta' \gamma , S^{MS'}\eta'\gamma, \left( {E\eta' \gamma } \right)\left[ P_1\eta\gamma\left\{ r'\eta\gamma /x \right\} \right], K'\gamma, \Lambda'\eta'\gamma } \right) \\
= &\left( {\underline{n}',S'\eta' \gamma , S^{MS'}\eta'\gamma, \mathcal {P}' \eta' \gamma, K'\gamma, \Lambda'\eta'\gamma } \right) \\
\end{align*}
\noindent Since we maintain the invariant that all bound variables
in $P_0$ are distinct from all other variables in $P_0$, $S$,
$\Lambda$, or $dom(\eta)$, we have $x\notin fv(E)\cup fv(S')\cup
fv(\Lambda)\cup dom(S^{MS}) \cup dom(\eta)$. Hence
$E\eta\gamma=E\eta'\gamma$, $S\eta \gamma=S'\eta' \gamma$,
$S^{MS}\eta \gamma=S^{MS'}\eta' \gamma$, and
$\Lambda\eta\gamma=\Lambda'\eta'\gamma$. Moreover, $P_1\eta \gamma
\left\{ {r'\eta\gamma /x} \right\} = {P_1}\left\{ {x: = r'}
\right\}\eta \gamma  = {P_1}\eta'\gamma $. Thus the last equation is
true.

ix) $\mathcal {P} = E[\text{lookup }M\text{ as }x\text{ in
}P_1\text{ else }P_2]$ with $t_M=eval^{CoSP}(M\eta\mu)$, and
$\forall (r,r')\in S$, $r\ncong t_M$. Then $K' = K,\mathcal
{P}'=E[P_2],\eta' = \eta ,\mu' = \mu, S' = S, S^{MS'} =
S^{MS},\Lambda' = \Lambda$. By (1), $ t_M\eta\gamma =
eval^{\pi}(M\eta\gamma) = M\eta\gamma$. $r\ncong t_M$ implies $r\eta
\ncong t_M\eta$. Since a Dolev-Yao adversary will never derive
protocol nonces that have never been sent, we have that only nonces
$N\in \textbf{N}_E \cup \text{range }\mu$ occur in $r\eta$ and
$t_M\eta$. By (3), $M\eta\gamma = t_M\eta\gamma\neq r\eta\gamma$ for
all $r\eta\gamma \in dom(S)\eta\gamma$. Thus we have
\begin{align*}
  &\left( {\underline{n},S\eta \gamma , S^{MS}\eta\gamma, \mathcal {P} \eta \gamma, K\gamma, \Lambda\eta\gamma } \right) \\
= &( \underline{n},S\eta \gamma , S^{MS}\eta\gamma,\left( {E\eta \gamma } \right)[ \text{lookup }M\eta\gamma\text{ as }x\text{ in }P_1\eta\gamma\text{ else }P_2\eta\gamma ], K\gamma, \Lambda\eta\gamma) \\
\to &\left( {\underline{n},S\eta \gamma ,S^{MS}\eta\gamma,\left( {E\eta \gamma } \right)\left[ P_2\eta\gamma \right], K\gamma , \Lambda\eta\gamma} \right) \\
= &\left( {\underline{n}',S'\eta' \gamma , S^{MS'}\eta'\gamma, \mathcal {P}' \eta' \gamma, K'\gamma, \Lambda'\eta'\gamma } \right) \\
\end{align*}

x) $\mathcal {P} = E[\text{lock }M;P_1]$ with
$t_M=eval^{CoSP}(M\eta\mu)$, and $\forall r\in \Lambda$, $r\ncong
t_M$. Then $K' = K,\mathcal {P}'=E[P_1],\eta' = \eta ,\mu' = \mu, S'
= S, S^{MS'} = S^{MS},\Lambda' = \Lambda\cup\{t_M\}$. By (1), $
t_M\eta\gamma = eval^{\pi}(M\eta\gamma) = M\eta\gamma$. $r\ncong
t_M$ implies $r\eta \ncong t_M\eta$. Since a Dolev-Yao adversary
will never derive protocol nonces that have never been sent, we have
that only nonces $N\in \textbf{N}_E \cup \text{range }\mu$ occur in
$r\eta$ and $t_M\eta$. By (3), $M\eta\gamma = t_M\eta\gamma\neq
r\eta\gamma$ for all $r\eta\gamma \in \Lambda\eta\gamma$. Thus we
have
\begin{align*}
  &\left( {\underline{n},S\eta \gamma , S^{MS}\eta\gamma, \mathcal {P} \eta \gamma, K\gamma, \Lambda\eta\gamma } \right) \\
= &\left( {\underline{n},S\eta \gamma , S^{MS}\eta\gamma,\left( {E\eta \gamma } \right)\left[ \text{lock }M\eta\gamma;P_1\eta\gamma \right], K\gamma, \Lambda\eta\gamma} \right) \\
\to &\left( {\underline{n},S\eta \gamma ,S^{MS}\eta\gamma,\left( {E\eta \gamma } \right)\left[ P_1\eta\gamma \right], K\gamma , \Lambda\eta\gamma \cup\{ M\eta\gamma \} } \right) \\
=   &\left( {\underline{n},S\eta \gamma ,S^{MS}\eta\gamma,\left( {E\eta \gamma } \right)\left[ P_1\eta\gamma \right], K\gamma , \Lambda\eta\gamma \cup\{ t_M\eta\gamma \} } \right) \\
= &\left( {\underline{n}',S'\eta' \gamma , S^{MS'}\eta'\gamma, \mathcal {P}' \eta' \gamma, K'\gamma, \Lambda'\eta'\gamma } \right) \\
\end{align*}

xi) $\mathcal {P} = E[\text{unlock }M;P_1]$ with
$t_M=eval^{CoSP}(M\eta\mu)$, and $\exists r\in \Lambda$ such that
$r\cong t_M$. Then $K' = K,\mathcal {P}'=E[P_1],\eta' = \eta ,\mu' =
\mu, S' = S, S^{MS'} = S^{MS},\Lambda' = \Lambda\backslash\{r\}$. By
(1), $ t_M\eta\gamma = eval^{\pi}(M\eta\gamma) = M\eta\gamma$.
$r\cong t_M$ implies $r\eta \cong t_M\eta$. Since a Dolev-Yao
adversary will never derive protocol nonces that have never been
sent, we have that only nonces $N\in \textbf{N}_E \cup \text{range
}\mu$ occur in $r\eta$ and $t_M\eta$. By (3), $M\eta\gamma =
t_M\eta\gamma= r\eta\gamma \in \Lambda\eta\gamma$. Thus we have
\begin{align*}
  &\left( {\underline{n},S\eta \gamma , S^{MS}\eta\gamma, \mathcal {P} \eta \gamma, K\gamma, \Lambda\eta\gamma } \right) \\
= &\left( {\underline{n},S\eta \gamma , S^{MS}\eta\gamma,\left( {E\eta \gamma } \right)\left[ \text{unlock }M\eta\gamma;P_1\eta\gamma \right], K\gamma, \Lambda\eta\gamma} \right) \\
\to &\left( {\underline{n},S\eta \gamma ,S^{MS}\eta\gamma,\left( {E\eta \gamma } \right)\left[ P_1\eta\gamma \right], K\gamma , \Lambda\eta\gamma \backslash\{ M\eta\gamma \} } \right) \\
=   &\left( {\underline{n},S\eta \gamma ,S^{MS}\eta\gamma,\left( {E\eta \gamma } \right)\left[ P_1\eta\gamma \right], K\gamma , \Lambda\eta\gamma \backslash\{ t_M\eta\gamma \} } \right) \\
= &\left( {\underline{n}',S'\eta' \gamma , S^{MS'}\eta'\gamma, \mathcal {P}' \eta' \gamma, K'\gamma, \Lambda'\eta'\gamma } \right) \\
\end{align*}

xii) $\mathcal {P} = E[[L]-[e]\rightarrow[R];P_1]$ with
$\tau':=\{x_i \mapsto s_i\}$ is a substitution found by the pattern
matching algorithm in $SExec_{P_0}^S$, and such that
$lfacts(L)(\eta\cup\tau')\mu \subseteq^\# {S^{MS}},
pfacts(L)(\eta\cup\tau')\mu \subset {S^{MS}}$. Then $K' = K,\mathcal
{P}'=E[P_1],\eta' = \eta\cup \tau' ,\mu' = \mu, S' = S, S^{MS'} =
S^{MS} \backslash^\# lfacts(L)\eta'\mu \cup^\# R\eta'\mu, \Lambda' =
\Lambda$. According to the pattern matching sub-protocol $f_{match}$, $\tau'$ is
grounding for $L$ such that for all symbolic fact $l\in^\#
lfacts(L)$, there exists a fact $f \in^\# S^{MS}$ such that
$l\eta'\mu = f$, where the equality means that the two symbolic
facts have the same fact label, and their arguments (as CoSP-terms)
are equal under $\cong$. Then for any $M_l$ that is the argument of
$l$, let $t_f$ be the corresponding one of $f$, by (1) we get
$M_l\eta'\gamma=eval^{CoSP}(M_l\eta'\mu)\eta'\gamma =
t_f\eta'\gamma$. Since all the arguments of $l\eta'\gamma$ and
$f\eta'\gamma$ are equal, we have that $l\eta'\gamma=f\eta'\gamma$.
Hence $lfacts(L\eta'\gamma) \subseteq^\# S^{MS}\eta'\gamma = S^{MS}
\eta\gamma$. It is similar to get $pfacts(L\eta'\gamma) \subset
S^{MS}\eta\gamma$. We set $\tau = \tau' \gamma$. Then we have
$(L\eta\gamma)\tau = (L\eta\gamma)(\tau'\gamma) = L\eta'\gamma =
(L\eta'\mu)\eta'\gamma$. The last equation is true since
$dom(\eta)\cap dom(\tau') = \emptyset$. Similarly, we have
$(R\eta\gamma)\tau=R\eta'\gamma = (R\eta'\mu)\eta'\gamma$. Thus we
have that
\begin{align*}
  &\left( {\underline{n},S\eta \gamma , S^{MS}\eta\gamma, \mathcal {P} \eta \gamma, K\gamma, \Lambda\eta\gamma } \right) \\
= &( \underline{n},S\eta \gamma , S^{MS}\eta'\gamma,\left( {E\eta \gamma } \right)\left[ ([L\eta\gamma]-[e]\rightarrow[R\eta\gamma]);P_1\eta\gamma \right], K\gamma, \Lambda\eta\gamma) \\
\to &( \underline{n},S\eta \gamma ,S^{MS}\eta'\gamma\backslash^\# lfacts(L\eta\gamma)\tau \cup^\# (R\eta\gamma)\tau ,\left( {E\eta \gamma } \right)\left[ (P_1\eta\gamma)\tau \right],K\gamma , \Lambda\eta\gamma  ) \\
=   &( \underline{n},S\eta \gamma ,S^{MS}\eta'\gamma\backslash^\# lfacts(L\eta'\mu)\eta'\gamma \cup^\# (R\eta'\mu)\eta'\gamma  ,\left( {E\eta \gamma } \right)\left[ (P_1\eta\gamma)\tau \right], K\gamma , \Lambda\eta\gamma ) \\
=   &( \underline{n},S\eta \gamma ,S^{MS'}\eta'\gamma,\left( {E\eta \gamma } \right)\left[ (P_1\eta\gamma)\tau \right], K\gamma , \Lambda\eta\gamma  ) \\
= &\left( {\underline{n}',S'\eta' \gamma , S^{MS'}\eta'\gamma, \mathcal {P}' \eta' \gamma, K'\gamma, \Lambda'\eta'\gamma } \right) \\
\end{align*}
\noindent Since we maintain the invariant that all bound variables
are pairwise distinct, $\forall x\in \tau$, we have $x\notin
fv(E)\cup fv(S')\cup fv(\Lambda')\cup fv(S^{MS}) \cup dom(\eta)$.
Hence $E\eta\gamma=E\eta'\gamma$, $S\eta \gamma=S'\eta' \gamma$, and
$\Lambda\eta\gamma=\Lambda'\eta'\gamma$. Moreover, we have
$(P_1\eta\gamma)\tau = (P_1\eta\gamma)(\tau'\gamma) =
P_1\eta'\gamma$. Thus the last equation is true.

xiii) In all other cases we have $\mathcal {P}'=\mathcal {P}, K' = K
,\eta' = \eta,\mu' = \mu, S' = S, S^{MS'} = S^{MS}, \Lambda' =
\Lambda$ and that $( {\underline{n},S\eta \gamma , S^{MS}\eta
\gamma, \mathcal {P} \eta \gamma, K\gamma, \Lambda\eta\gamma }
)= ( \underline{n}',\eta' \gamma , \\ S^{MS'}\eta \gamma,
\mathcal {P}' \eta' \gamma, K'\gamma, \Lambda'\eta'\gamma )$
\end{proof}

\noindent\textbf{Theorem 1 (CS in SAPIC)}. Assume that the
computational implementation of the applied $\pi$-calculus is a
computationally sound implementation (Definition 3) of the symbolic
model of the applied $\pi$-calculus (Definition 4) for a class
\textbf{P} of protocols. If a closed SAPIC process $P_0$
symbolically satisfies a SAPIC trace property $\wp$ (Definition 7),
and $\Pi_{P_0}^S\in\textbf{P}$, then $P_0$ computationally satisfies
$\wp$ (Definition 12).

\begin{proof}. Assume that $P_0$ symbolically satisfies $\wp$. By
lemma 3, $SExec_{P_0}^S$ satisfies $\wp$. By lemma 1, $\Pi_{P_0}^S$
symbolically satisfies $events^{-1}(\wp)$. Furthermore, since $\wp$
is an efficiently decidable, prefix closed set, so is
$events^{-1}(\wp)$. Thus $events^{-1}(\wp)$ is a CoSP-trace
property. By lemma 2, we have that $\Pi_{P_0}^{S}$ is an efficient
CoSP protocol. By assumption, the computational implementation $A$
of the applied $\pi$-calculus is computationally sound; hence
$(\Pi_{P_0}^{S}, A)$ computationally satisfies $events^{-1}(\wp)$.
Using lemma 1, we obtain that $P_0$ computationally satisfies $\wp$.
\end{proof}
\newpage

\section*{Appendix B: Proof of Theorem 2}

\noindent\textbf{Lemma 4}. Let $O_1$ be a StatVerif semantic
configuration. If $O_1 \xrightarrow{\alpha} O_2$, then $\lfloor
O_1\rfloor \xrightarrow{\alpha}^* \lfloor O_2\rfloor$.

\begin{proof}. We prove this lemma by induction over the size of the
set of processes in $O_1$. Let $O_1 = ( \tilde
n,\mathcal{S},\mathcal{P} \cup \{ ( P_0,\beta_0  ) \}, \mathcal{K}
)$ be a StatVerif semantic configuration, where $\mathcal{P}=
\mathop  \cup \limits_{i = 1}^k \{ ( P_i,\beta _i ) \}$. Assume that
$O_1 \xrightarrow{\alpha} O_2$ conducts a reduction on $(P_0,
\beta_0)$. We distinguish the following cases of $P_0$:

i) In the following cases, where $P_0 = P|0$, or $P_0=!P$, or $P_0 = P|Q$, or
$P_0 = \nu n;P$, or $P_0 = \text{let }x=D\text{ in }P\text{ else }Q$, or
$P_0 = \text{out}(M,N);P$, or $P_0=\text{in}(M,x);P$, or $P_0=\text{event }e;P$,
we have that $\lfloor P_0 \rfloor_{\beta_0}$ keep the standard constructs
unchanged. Thus it is easy to obtain $\lfloor
O_1\rfloor \xrightarrow{\alpha}^* \lfloor O_2\rfloor$ where $O_1 \xrightarrow{\alpha} O_2$
conducts a reduction on $(P_0, \beta_0)$.

ii) $O_1 = (\tilde n, \mathcal{S}, \mathcal{P}\cup \{([s\mapsto M], 0)\}, \mathcal{K})$,
$O_2 = (\tilde n, \mathcal{S}\cup\{s:=M\}, \mathcal{P}, \mathcal{K})$, and $s\in\tilde{n},s\notin dom(\mathcal{S})$.
Then we have

\begin{align*}
  \lfloor O_1 \rfloor = & (\tilde{n}, \mathcal{S}, \emptyset, \mathop\cup\limits_{i = 1}^k\{\lfloor P_i \rfloor_{\beta_i}\}\cup\{\text{insert }s,M\}, \mathcal{K}, \Lambda ) \\
\to & (\tilde{n}, \mathcal{S}\cup\{s:=M\}, \emptyset, \mathop\cup\limits_{i = 1}^k\{\lfloor P_i \rfloor_{\beta_i}\}, \mathcal{K}, \Lambda)\\
= & \lfloor O_2 \rfloor \\
\end{align*}

\noindent where $\Lambda = \{l\}$ if $\exists (P_i, \beta_i)\in \mathcal P, \beta_i =1$, or
$\Lambda = \emptyset$ if $\forall (P_i, \beta_i)\in \mathcal P, \beta_i =0$.

iii) $O_1 = (\tilde n, \mathcal{S}, \mathcal{P}\cup \{(s:=M;P, 0)\}, \mathcal{K})$,
$O_2 = (\tilde n, \mathcal{S}\cup\{s:=M\}, \mathcal{P}\cup \{(P,0)\}, \mathcal{K})$, and $s\in dom(\mathcal{S}), \forall (P_i, \beta_i)\in \mathcal P, \beta_i=0$.
Then we have

\begin{align*}
  \lfloor O_1 \rfloor = & (\tilde{n}, \mathcal{S}, \emptyset, \mathop\cup\limits_{i = 1}^k\{\lfloor P_i \rfloor_{\beta_i}\}\cup\{\text{lock }l;\text{lookup }s\text{ as }x_s\text{ in insert }s,M;\text{unlock }l;\lfloor P\rfloor_0\}, \mathcal{K}, \emptyset ) \\
\to & (\tilde{n}, \mathcal{S}, \emptyset, \mathop\cup\limits_{i = 1}^k\{\lfloor P_i \rfloor_{\beta_i}\}\cup\{ \text{lookup }s\text{ as }x_s\text{ in insert }s,M;\text{unlock }l;\lfloor P\rfloor_0\}, \mathcal{K}, \{l\} )\\
\to & (\tilde{n}, \mathcal{S}, \emptyset, \mathop\cup\limits_{i = 1}^k\{\lfloor P_i \rfloor_{\beta_i}\}\cup\{ \text{insert }s,M;\text{unlock }l;\lfloor P\rfloor_0\}, \mathcal{K}, \{l\} )\\
\to & (\tilde{n}, \mathcal{S}\cup\{s:=M\}, \emptyset, \mathop\cup\limits_{i = 1}^k\{\lfloor P_i \rfloor_{\beta_i}\}\cup\{ \text{unlock }l;\lfloor P\rfloor_0\}, \mathcal{K}, \{l\} )\\
\to & (\tilde{n}, \mathcal{S}\cup\{s:=M\}, \emptyset, \mathop\cup\limits_{i = 1}^k\{\lfloor P_i \rfloor_{\beta_i}\}\cup\{ \lfloor P\rfloor_0\}, \mathcal{K}, \emptyset )\\
= & \lfloor O_2 \rfloor \\
\end{align*}

\noindent Note that the second reduction is true because $x_s$ is fresh.

iv) $O_1 = (\tilde n, \mathcal{S}, \mathcal{P}\cup \{(s:=M;P, 1)\}, \mathcal{K})$,
$O_2 = (\tilde n, \mathcal{S}\cup\{s:=M\}, \mathcal{P}\cup \{(P,1)\}, \mathcal{K})$, and $s\in dom(\mathcal{S}), \forall (P_i, \beta_i)\in \mathcal P, \beta_i=0$.
Then we have

\begin{align*}
  \lfloor O_1 \rfloor = & (\tilde{n}, \mathcal{S}, \emptyset, \mathop\cup\limits_{i = 1}^k\{\lfloor P_i \rfloor_{\beta_i}\}\cup\{\text{lookup }s\text{ as }x_s\text{ in insert }s,M;\lfloor P\rfloor_1\}, \mathcal{K},\{l\}) \\
\to & (\tilde{n}, \mathcal{S}, \emptyset, \mathop\cup\limits_{i = 1}^k\{\lfloor P_i \rfloor_{\beta_i}\}\cup\{ \text{insert }s,M;\lfloor P\rfloor_1\}, \mathcal{K}, \{l\} )\\
\to & (\tilde{n}, \mathcal{S}\cup\{s:=M\}, \emptyset, \mathop\cup\limits_{i = 1}^k\{\lfloor P_i \rfloor_{\beta_i}\}\cup\{ \lfloor P\rfloor_1\}, \mathcal{K}, \{l\} )\\
= & \lfloor O_2 \rfloor \\
\end{align*}

\noindent Note that the first reduction is true because $x_s$ is fresh.

v) $O_1 = (\tilde n, \mathcal{S}, \mathcal{P}\cup \{(\text{read }s \text{ as }x;P, 0)\}, \mathcal{K})$,
$O_2 = (\tilde n, \mathcal{S}, \mathcal{P}\cup \{(P\{\mathcal S(s)/x\},0)\}, \mathcal{K})$, and $s\in dom(\mathcal{S}), \forall (P_i, \beta_i)\in \mathcal P, \beta_i=0$.
Then we have

\begin{align*}
  \lfloor O_1 \rfloor = & (\tilde{n}, \mathcal{S}, \emptyset, \mathop\cup\limits_{i = 1}^k\{\lfloor P_i \rfloor_{\beta_i}\}\cup\{\text{lock }l;\text{lookup }s\text{ as }x\text{ in unlock }l;\lfloor P\rfloor_0\}, \mathcal{K}, \emptyset ) \\
\to & (\tilde{n}, \mathcal{S}, \emptyset, \mathop\cup\limits_{i = 1}^k\{\lfloor P_i \rfloor_{\beta_i}\}\cup\{ \text{lookup }s\text{ as }x\text{ in unlock }l;\lfloor P\rfloor_0\}, \mathcal{K}, \{l\} )\\
\to & (\tilde{n}, \mathcal{S}, \emptyset, \mathop\cup\limits_{i = 1}^k\{\lfloor P_i \rfloor_{\beta_i}\}\cup\{ \text{unlock }l;\lfloor P\{\mathcal S(s)/x\}\rfloor_0\}, \mathcal{K}, \{l\} )\\
\to & (\tilde{n}, \mathcal{S}, \emptyset, \mathop\cup\limits_{i = 1}^k\{\lfloor P_i \rfloor_{\beta_i}\}\cup\{ \lfloor P\{\mathcal S(s)/x\}\rfloor_0\} , \mathcal{K}, \emptyset )\\
= & \lfloor O_2 \rfloor \\
\end{align*}

vi) $O_1 = (\tilde n, \mathcal{S}, \mathcal{P}\cup \{(\text{read }s \text{ as }x;P, 1)\}, \mathcal{K})$,
$O_2 = (\tilde n, \mathcal{S}, \mathcal{P}\cup \{(P\{\mathcal S(s)/x\},1)\}, \mathcal{K})$, and $s\in dom(\mathcal{S}), \forall (P_i, \beta_i)\in \mathcal P, \beta_i=0$.
Then we have

\begin{align*}
  \lfloor O_1 \rfloor = & (\tilde{n}, \mathcal{S}, \emptyset, \mathop\cup\limits_{i = 1}^k\{\lfloor P_i \rfloor_{\beta_i}\}\cup\{\text{lookup }s\text{ as }x\text{ in }\lfloor P\rfloor_1\}, \mathcal{K}, \{l\} ) \\
\to & (\tilde{n}, \mathcal{S}, \emptyset, \mathop\cup\limits_{i = 1}^k\{\lfloor P_i \rfloor_{\beta_i}\}\cup\{ \lfloor P\{\mathcal S(s)/x\}\rfloor_1\} , \mathcal{K}, \{l\} )\\
= & \lfloor O_2 \rfloor \\
\end{align*}

vii) $O_1 = (\tilde n, \mathcal{S}, \mathcal{P}\cup \{(\text{lock};P, 0)\}, \mathcal{K})$,
$O_2 = (\tilde n, \mathcal{S}, \mathcal{P}\cup \{(P,1)\}, \mathcal{K})$, and $\forall (P_i, \beta_i)\in \mathcal P, \beta_i=0$.
Then we have

\begin{align*}
  \lfloor O_1 \rfloor = & (\tilde{n}, \mathcal{S}, \emptyset, \mathop\cup\limits_{i = 1}^k\{\lfloor P_i \rfloor_{\beta_i}\}\cup\{\text{lock }l;\lfloor P\rfloor_1\}, \mathcal{K}, \emptyset ) \\
\to & (\tilde{n}, \mathcal{S}, \emptyset, \mathop\cup\limits_{i = 1}^k\{\lfloor P_i \rfloor_{\beta_i}\}\cup\{ \lfloor P\rfloor_1\} , \mathcal{K}, \{l\} )\\
= & \lfloor O_2 \rfloor \\
\end{align*}

viii) $O_1 = (\tilde n, \mathcal{S}, \mathcal{P}\cup \{(\text{unlock};P, 1)\}, \mathcal{K})$,
$O_2 = (\tilde n, \mathcal{S}, \mathcal{P}\cup \{(P,0)\}, \mathcal{K})$, and $\forall (P_i, \beta_i)\in \mathcal P, \beta_i=0$.
Then we have

\begin{align*}
  \lfloor O_1 \rfloor = & (\tilde{n}, \mathcal{S}, \emptyset, \mathop\cup\limits_{i = 1}^k\{\lfloor P_i \rfloor_{\beta_i}\}\cup\{\text{unlock }l;\lfloor P\rfloor_0\}, \mathcal{K}, \{l\} ) \\
\to & (\tilde{n}, \mathcal{S}, \emptyset, \mathop\cup\limits_{i = 1}^k\{\lfloor P_i \rfloor_{\beta_i}\}\cup\{ \lfloor P\rfloor_0\} , \mathcal{K}, \emptyset )\\
= & \lfloor O_2 \rfloor \\
\end{align*}

ix) In all the other cases, there is no reduction for $O_1 \xrightarrow{\alpha} O_2$ that conducts a reduction on $(P_0, \beta_0)$.
\end{proof}

\noindent\textbf{Lemma 5}. Let $O_1$ be a StatVerif semantic
configuration. If $\lfloor O_1\rfloor \xrightarrow{\alpha} O'$, then
there exists a StatVerif semantic configuration $O_2$, such that
$O_1 \xrightarrow{\alpha}^*  O_2$ and that $O' = \lfloor O_2\rfloor$
or $O' \xrightarrow{}^* \lfloor O_2\rfloor$.

\begin{proof}.
We prove this lemma by induction over the size of the
set of processes in $\lfloor O_1\rfloor$. Let $\lfloor O_1\rfloor =
( \tilde n,\mathcal{S}, \emptyset,\mathcal{P}'  \cup \{ P_0 \},
\mathcal{K}, \Lambda )$ be a SAPIC semantic configuration
transformed from a StatVerif semantic configuration, where
$\mathcal{P}'= \mathop  \cup \limits_{i = 1}^k \{ P_i'\}$. Assume
that $\lfloor O_1\rfloor \xrightarrow{\alpha} O'$ conducts a
reduction on $P_0$. We distinguish the following cases of $P_0$:

i) In the following cases, where $P_0 = P|0$, or $P_0=!P$, or $P_0 = P|Q$, or
$P_0 = \nu n;P$, or $P_0 = \text{let }x=D\text{ in }P\text{ else }Q$, or
$P_0 = \text{out}(M,N);P$, or $P_0=\text{in}(M,x);P$, or $P_0=\text{event }e;P$,
we have that $\lfloor \cdot \rfloor_{\beta}$ keep the standard constructs
unchanged. Thus it is easy to obtain the StatVerif semantic configuration $O_2$ such that $O_1 \xrightarrow{\alpha} O_2$ and $O' = \lfloor O_2\rfloor$.

ii) $\lfloor O_1\rfloor = (\tilde n, \mathcal{S}, \emptyset, \mathcal{P}'\cup \{\text{insert }s,M;P'\}, \mathcal{K}, \Lambda)$,
$O' = (\tilde n, \mathcal{S}\cup\{s:=M\}, \emptyset, \mathcal{P}'\cup\{P'\}, \mathcal{K}, \Lambda)$. We get $\lfloor O_1\rfloor \xrightarrow{\alpha} O'$. According to the rules of encoding, we can assume $O_1 = (\tilde n, \mathcal S, \mathcal P \cup \{([s\mapsto M],0)\}, \mathcal K)$. Let $O_2 = (\tilde n, \mathcal S \cup\{s:=M\}, \mathcal P , \mathcal K)$ be a StatVerif semantic configuration, we have $O' = \lfloor O_2\rfloor$. It is left to show $O_1 \xrightarrow{\alpha} O_2$. This reduction needs two conditions: $s\in \tilde n$ and $s\notin dom(\mathcal S)$. We get $s\in \tilde n$ from the fact that $[s\mapsto M]$ is a process in $O_1$ and from the first restriction in the syntax of StatVerif. For $s\notin dom(\mathcal S)$, we use the disproof method. If $s\in dom(\mathcal S)$, the first insertion for the state cell $s$ should be performed by the process $\lfloor [s\mapsto N]\rfloor_0$ or $\lfloor s:=N;P\rfloor_{\beta}$. The former contradicts the restriction that $[s\mapsto N]$ occurs only once. The latter cannot be the first time to perform the insertion since we set a lookup construct before the insert construct. Thus $s\notin dom(\mathcal S)$ and we have $O_1 \xrightarrow{\alpha} O_2$.

iii) $\lfloor O_1\rfloor = (\tilde n, \mathcal{S}, \emptyset, \mathcal{P}'\cup \{\text{unlock }l;P'\}, \mathcal{K}, \{l\})$,
$O' = (\tilde n, \mathcal{S}, \emptyset, \mathcal{P}'\cup\{P'\}, \mathcal{K}, \emptyset)$. We get $\lfloor O_1\rfloor \xrightarrow{\alpha} O'$. According to the rules of encoding, we can assume $O_1 = (\tilde n, \mathcal S, \mathcal P \cup \{(\text{unlock};P, 1)\}, \mathcal K)$. Let $O_2 = (\tilde n, \mathcal S , \mathcal P \cup\{(P,0)\} , \mathcal K)$ be a StatVerif semantic configuration, we have $O' = \lfloor O_2\rfloor$ and $O_1 \xrightarrow{\alpha} O_2$.

iv) $\lfloor O_1\rfloor = (\tilde n, \mathcal{S}, \emptyset, \mathcal{P}'\cup \{\text{lock }l;P'\}, \mathcal{K}, \emptyset)$, $O' = (\tilde n, \mathcal{S}, \emptyset, \mathcal{P}'\cup \{P'\}, \mathcal{K}, \{l\})$. We get $\lfloor O_1\rfloor \xrightarrow{\alpha} O'$. According to the rules of encoding, we distinguish 3 cases in the construction of $O_1$:

\noindent (a) We assume $O_1 = (\tilde n, \mathcal S, \mathcal P \cup \{(\text{lock};P, 0)\}, \mathcal K)$ and $\forall (P_i, \beta_i)\in \mathcal P, \beta_i=0$. Let $O_2 = (\tilde n, \mathcal S , \mathcal P \cup\{(P,1)\} , \mathcal K)$ be a StatVerif semantic configuration, we have $O' = \lfloor O_2\rfloor$ and $O_1 \xrightarrow{\alpha} O_2$.

\noindent (b) We assume $O_1 = (\tilde n, \mathcal S, \mathcal P \cup \{(s:=M;P, 0)\}, \mathcal K)$ and $\forall (P_i, \beta_i)\in \mathcal P, \beta_i=0$. According to the rules of encoding, we have $P' = \text{lookup }s\text{ as }x_s \text{ in insert }s,M;\text{unlock }l;\lfloor P\rfloor_0$. If $\mathcal P = \{([s\mapsto N],0)\}\cup\mathcal P_1$, then set $O_2 = (\tilde n, \mathcal S \cup\{s:=M\}, \mathcal P_1 \cup\{(P,0)\} , \mathcal K)$. Otherwise, set $O_2 = (\tilde n, \mathcal S \cup\{s:=M\}, \mathcal P \cup\{(P,0)\} , \mathcal K)$. For $\mathcal P = \{([s\mapsto N],0)\}\cup\mathcal P_1$, we have $O_1 \xrightarrow{}^* O_2$. It is left to show $O' \xrightarrow{}^* \lfloor O_2\rfloor$. We can assume $\lfloor O_2\rfloor = (\tilde n, \mathcal S \cup\{s:=M\}, \emptyset, \mathcal P_1' \cup\{\lfloor P\rfloor_0\} , \mathcal K, \empty)$ where $\mathcal P_1' = \mathcal P' \backslash \{\text{insert }s,N\}$. Then we have

\begin{align*}
  O'  = & (\tilde{n}, \mathcal{S}, \emptyset, \mathcal P_1'\cup\{\text{insert }s,N\} \cup\{ \text{lookup }s\text{ as }x_s\text{ in insert }s,M;\text{unlock }l;\lfloor P\rfloor_0\}, \mathcal{K}, \{l\} )\\
\to & (\tilde{n}, \mathcal{S}\cup\{s:=N\}, \emptyset, \mathcal P_1'\cup\{ \text{lookup }s\text{ as }x_s\text{ in insert }s,M;\text{unlock }l;\lfloor P\rfloor_0\}, \mathcal{K}, \{l\} )\\
\to & (\tilde{n}, \mathcal{S}\cup\{s:=N\}, \emptyset, \mathcal P_1'\cup\{ \text{insert }s,M;\text{unlock }l;\lfloor P\rfloor_0\}, \mathcal{K}, \{l\} )\\
\to & (\tilde{n}, \mathcal{S}\cup\{s:=M\}, \emptyset, \mathcal P_1'\cup\{ \text{unlock }l;\lfloor P\rfloor_0\}, \mathcal{K}, \{l\} )\\
\to & (\tilde{n}, \mathcal{S}\cup\{s:=M\}, \emptyset, \mathcal P_1'\cup\{ \lfloor P\rfloor_0\}, \mathcal{K}, \emptyset )\\
= & \lfloor O_2 \rfloor \\
\end{align*}

\noindent For $\{([s\mapsto N],0)\}\notin \mathcal P $, we get $s\in dom(\mathcal S)$ from the restriction of the syntax of $[s\mapsto M]$. Thus we have that $O_1 \xrightarrow{}^* O_2$, and that

\begin{align*}
  O'  = & (\tilde{n}, \mathcal{S}, \emptyset, \mathcal P' \cup\{ \text{lookup }s\text{ as }x_s\text{ in insert }s,M;\text{unlock }l;\lfloor P\rfloor_0\}, \mathcal{K}, \{l\} )\\
\to & (\tilde{n}, \mathcal{S}, \emptyset, \mathcal P'\cup\{ \text{insert }s,M;\text{unlock }l;\lfloor P\rfloor_0\}, \mathcal{K}, \{l\} )\\
\to & (\tilde{n}, \mathcal{S}\cup\{s:=M\}, \emptyset, \mathcal P'\cup\{ \text{unlock }l;\lfloor P\rfloor_0\}, \mathcal{K}, \{l\} )\\
\to & (\tilde{n}, \mathcal{S}\cup\{s:=M\}, \emptyset, \mathcal P'\cup\{ \lfloor P\rfloor_0\}, \mathcal{K}, \emptyset )\\
= & \lfloor O_2 \rfloor \\
\end{align*}

\noindent (c) We assume $O_1 = (\tilde n, \mathcal S, \mathcal P \cup \{(\text{read }s\text{ as }x;P, 0)\}, \mathcal K)$ and $\forall (P_i, \beta_i)\in \mathcal P, \beta_i=0$. According to the rules of encoding, we have $P' = \text{lookup }s\text{ as }x \text{ in unlock }l;\lfloor P\rfloor_0$. If $\mathcal P = \{([s\mapsto N],0)\}\cup\mathcal P_1$, then set $O_2 = (\tilde n, \mathcal S, \mathcal P_1 \cup\{(P\{N/x\},0)\} , \mathcal K)$. Otherwise, set $O_2 = (\tilde n, \mathcal S , \mathcal P \cup\{(P\{\mathcal S(s)/x\},0)\} , \mathcal K)$. For $\mathcal P = \{([s\mapsto N],0)\}\cup\mathcal P_1$, we have $O_1 \xrightarrow{}^* O_2$. It is left to show $O' \xrightarrow{}^* \lfloor O_2\rfloor$. We can assume $\lfloor O_2\rfloor = (\tilde n, \mathcal S , \emptyset, \mathcal P_1' \cup\{\lfloor P\{N/x \} \rfloor_0\} , \mathcal K, \emptyset)$ where $\mathcal P_1' = \mathcal P' \backslash \{\text{insert }s,N\}$. Then we have

\begin{align*}
  O'  = & (\tilde{n}, \mathcal{S}, \emptyset, \mathcal P_1'\cup\{\text{insert }s,N\} \cup\{ \text{lookup }s\text{ as }x\text{ in unlock }l;\lfloor P\rfloor_0\}, \mathcal{K}, \{l\} )\\
\to & (\tilde{n}, \mathcal{S}\cup\{s:=N\}, \emptyset, \mathcal P_1'\cup\{ \text{lookup }s\text{ as }x\text{ in unlock }l;\lfloor P\rfloor_0\}, \mathcal{K}, \{l\} )\\
\to & (\tilde{n}, \mathcal{S}\cup\{s:=N\}, \emptyset, \mathcal P_1'\cup\{ \text{unlock }l;\lfloor P\{N/x\}\rfloor_0\}, \mathcal{K}, \{l\} )\\
\to & (\tilde{n}, \mathcal{S}\cup\{s:=N\}, \emptyset, \mathcal P_1'\cup\{ \lfloor P\{N/x\}\rfloor_0\}, \mathcal{K}, \emptyset )\\
= & \lfloor O_2 \rfloor \\
\end{align*}

\noindent For $\{([s\mapsto N],0)\}\notin \mathcal P $, we get $s\in dom(\mathcal S)$ from the restriction of the syntax of $[s\mapsto M]$. Then we have that $O_1 \xrightarrow{}^* O_2$, and that

\begin{align*}
  O'  = & (\tilde{n}, \mathcal{S}, \emptyset, \mathcal P' \cup\{ \text{lookup }s\text{ as }x\text{ in unlock }l;\lfloor P\rfloor_0\}, \mathcal{K}, \{l\} )\\
\to & (\tilde{n}, \mathcal{S}, \emptyset, \mathcal P'\cup\{ \text{unlock }l;\lfloor P\{\mathcal S(s)/x \} \rfloor_0\}, \mathcal{K}, \{l\} )\\
\to & (\tilde{n}, \mathcal{S}, \emptyset, \mathcal P'\cup\{ \lfloor P\{\mathcal S(s)/x \} \rfloor_0\}, \mathcal{K}, \emptyset )\\
= & \lfloor O_2 \rfloor \\
\end{align*}

v) $\lfloor O_1\rfloor = (\tilde n, \mathcal{S}, \emptyset, \mathcal{P}'\cup \{\text{lookup }s\text{ as }x\text{ in }P'\}, \mathcal{K}, \{l\})$, $O' = (\tilde n, \mathcal{S}, \emptyset, \mathcal{P}'\cup \{P'\{\mathcal S(s)/x\}\}, \mathcal{K}, \{l\})$. We get $\lfloor O_1\rfloor \xrightarrow{\alpha} O'$. According to the rules of encoding, we distinguish 2 cases in the construction of $O_1$:

\noindent (a) We assume $O_1 = (\tilde n, \mathcal S, \mathcal P \cup \{(s:=M;P, 1)\}, \mathcal K)$ and $\forall (P_i, \beta_i)\in \mathcal P, \beta_i=0$. According to the rules of encoding, we have $P' = \text{ insert }s,M;\lfloor P\rfloor_1$ and $x$ is a fresh variable. Let $O_2 = (\tilde n, \mathcal S \cup\{s:=M\}, \mathcal P \cup\{(P,1)\}, \mathcal K)$ be a StatVerif semantic configuration. Since $\lfloor O_1 \rfloor \xrightarrow{} O'$ conducts a reduction on the lookup construct. We get $s\in dom(\mathcal S)$. Thus we have that $O_1 \xrightarrow{}^* O_2$, and that

\begin{align*}
  O'  = & (\tilde{n}, \mathcal{S}, \emptyset, \mathcal P'\cup\{P'\{\mathcal S(s)/x\}\}, \mathcal{K}, \{l\} )\\
= & (\tilde{n}, \mathcal{S}, \emptyset, \mathcal P'\cup\{ \text{insert }s,M;\lfloor P\rfloor_1\}, \mathcal{K}, \{l\} )\\
\to & (\tilde{n}, \mathcal{S}\cup\{s:=M\}, \emptyset, \mathcal P'\cup\{ \lfloor P\rfloor_1\}, \mathcal{K}, \{l\} )\\
= & \lfloor O_2 \rfloor \\
\end{align*}

\noindent (b) We assume $O_1 = (\tilde n, \mathcal S, \mathcal P \cup \{(\text{read }s\text{ as }x;P, 1)\}, \mathcal K)$ and $\forall (P_i, \beta_i)\in \mathcal P, \beta_i=0$. According to the rules of encoding, we have $P' = \lfloor P\rfloor_1$. Let $O_2 = (\tilde n, \mathcal S , \mathcal P \cup\{(P\{\mathcal S(s)/x\},1)\}, \mathcal K)$ be a StatVerif semantic configuration. Since $\lfloor O_1 \rfloor \xrightarrow{} O'$ conducts a reduction on the lookup construct. We get $s\in dom(\mathcal S)$. Thus we have that $O_1 \xrightarrow{}^* O_2$, and that

\begin{align*}
  O'  = & (\tilde{n}, \mathcal{S}, \emptyset, \mathcal P'\cup\{P'\{\mathcal S(s)/x\}\}, \mathcal{K}, \{l\} )\\
= & (\tilde{n}, \mathcal{S}, \emptyset, \mathcal P'\cup\{ \lfloor P\{\mathcal S(s)/x\}\rfloor_1\}, \mathcal{K}, \{l\} )\\
= & \lfloor O_2 \rfloor \\
\end{align*}

vi) In all the other cases, there is no reduction for $\lfloor O_1\rfloor \xrightarrow{\alpha} O'$ that conducts a reduction on $P_0$
\end{proof}

\noindent\textbf{Lemma 6}. Let $P_0$ be a closed StatVerif process.
Let $M$ be a message. Set $P':=\text{in}(attch, x);\text{let
}y=equal(x,M)$ in event $NotSecret $, where $x,y$ are two fresh
variables that are not used in $P_0$, $attch \in \textbf{N}_E$ is a
free channel name which is known by the adversary. We set $\wp
:=\{e|NotSecret$ is not in $e\}$. $Q_0:=\lfloor P'|P_0\rfloor_0$ is
a closed SAPIC process and $\wp $ is a SAPIC trace property. Then we
have that $P_0$ symbolically preserves the secrecy of $M$ (in the
sense of Definition 13) \emph{iff} $Q_0$ symbolically satisfies $\wp
$ (in the sense of Definition 7).

\begin{proof}.
By definition 13, $P_0$ does not preserve the secrecy
of $M$ if there exists a StatVerif trace of the form $
(\emptyset,\emptyset, \{(P_0,0)\}, fn(P_0))\xrightarrow{{\alpha}}^*
(\tilde{n}, \mathcal {S}, \mathcal {P}, \mathcal {K})$ where $\nu
\tilde{n}.\mathcal {K} \vdash M$. Then we have the following
StatVerif trace

\begin{align*}
  O&=(\emptyset,\emptyset, \{(Q_0,0)\}, fn(Q_0))\xrightarrow{} (\emptyset,\emptyset, \{(P_0,0)\}\cup\{(P',0)\}, fn(P_0)) \xrightarrow{\alpha}^* (\tilde{n}, \mathcal {S}, \mathcal {P}\cup \{(P',0)\}, \mathcal {K}) \\
&\xrightarrow{K(attch, M)}^* (\tilde{n}, \mathcal {S}, \mathcal {P} \cup\{\text{event }NotSecret\}, \mathcal {K}) \xrightarrow{NotSecret} (\tilde{n}, \mathcal {S}, \mathcal {P}, \mathcal {K}) \\
\end{align*}

\noindent By lemma 4, for $\lfloor O\rfloor$ there exists a trace
that contains the event $NotSecret$. Thus $Q_0$ does not satisfy $\wp$.

For the opposite direction, if $Q_0$ does not satisfy $\wp$, then we get
$(\emptyset, \emptyset, \emptyset, \{Q_0\}, \\fn(P_0), \emptyset) \xrightarrow{\alpha}^*
\xrightarrow{NotSecret} (\tilde{n}, \mathcal {S}, \emptyset, \mathcal {P}, \mathcal {K}, \Lambda)$.
We distinguish two cases for the reduction of in$(attch,x)$ construct in $P'$:

i) The adversary inputs a term $N$ on the channel $attch$. We have the following trace

\begin{align*}
  \lfloor O\rfloor&=(\emptyset, \emptyset, \emptyset, \{Q_0\}, fn(P_0), \emptyset)\xrightarrow{\alpha}^* (\tilde{n}_1, \mathcal{S}_1, \emptyset, \mathcal{P}_1 \cup\{P'\}, \mathcal {K}_1, \Lambda_1) \\ &\xrightarrow{K(attch, N)} (\tilde{n}_1, \mathcal {S}_1, \emptyset, \mathcal {P}_1 \cup\{\text{let }y=equal(N,M)\text{ in event }NotSecret\}, \mathcal {K}_1, \Lambda_1) \\
&\xrightarrow{NotSecret}^* (\tilde{n}, \mathcal {S}, \emptyset, \mathcal {P}, \mathcal {K}, \Lambda) \\
\end{align*}

\noindent We get that $M=_E N$. By lemma 5 and the first
reduction step, we have that $P_0$ does not preserve the
secrecy of $M$.

ii) Before the reduction of $P'$, the process has output a term
$N$ on the channel $attch$. We have the following trace.

\begin{align*}
  \lfloor O\rfloor&=(\emptyset, \emptyset, \emptyset, \{Q_0\}, fn(P_0), \emptyset)\\
  &\xrightarrow{\alpha}^* (\tilde{n}_1, \mathcal{S}_1, \emptyset, \mathcal{P}_1\cup\{\text{out}(attch,N);P_1\} \cup\{P'\}, \mathcal {K}_1, \Lambda_1) \\
  &\xrightarrow{} (\tilde{n}_1, \mathcal {S}_1, \emptyset, \mathcal {P}_1\cup\{P_1\} \cup\{\text{let }y=equal(N,M)\text{ in event }NotSecret\}, \mathcal {K}_1, \Lambda_1) \\
  &\xrightarrow{NotSecret}^* (\tilde{n}, \mathcal {S}, \emptyset, \mathcal {P}, \mathcal {K}, \Lambda) \\
\end{align*}

\noindent We have that $M=_E N$, and that

\begin{align*}
  \lfloor O\rfloor&=(\emptyset, \emptyset, \emptyset, \{Q_0\}, fn(P_0), \emptyset)\\
  &\xrightarrow{\alpha}^* (\tilde{n}_1, \mathcal{S}_1, \emptyset, \mathcal{P}_1\cup\{\text{out}(attch,N);P_1\} \cup\{P'\}, \mathcal {K}_1, \Lambda_1) \\
  &\xrightarrow{K(N)} (\tilde{n}_1, \mathcal {S}_1, \emptyset, \mathcal {P}_1\cup\{P_1\}\cup\{P'\}, \mathcal {K}_1\cup\{N\}, \Lambda_1) \\
\end{align*}

\noindent By lemma 5, we have that $P_0$ does not preserve the
secrecy of $M$.
\end{proof}

\noindent\textbf{Theorem 2 (CS in StatVerif)}. Assume that the
computational implementation of the applied $\pi$-calculus
is a computationally sound implementation (Definition 3)
of the symbolic model of the applied
$\pi$-calculus (Definition 4) for a class \textbf{P} of protocols.
For a closed StatVerif process $P_0$, we denote by $Q_0$ and $\wp$
the same meanings in Lemma 6. Thus if the StatVerif process $P_0$
symbolically preserves the secrecy of a message $M$ (Definition 13)
and $\Pi_{Q_0}^S\in\textbf{P}$, then $Q_0$ computationally satisfies $\wp$.

\begin{proof}. Theorem 2 can be easily proved by using Lemma 6 and
Theorem 1.
\end{proof}

\newpage

\section*{Appendix C: Left-or-right Protocol in StatVerif Syntax}

\begin{lstlisting}[language=C]
fun enc/3. fun ek/1. fun dk/1. fun sig/3. fun vk/1. fun sk/1.
fun pair/2. fun garbage/1. fun garbageEnc/2. fun garbageSig/2.
fun string0/1. fun string1/1. fun empty/0.
reduc dec(dk(t1),enc(ek(t1),m,t2)) = m.
reduc isek(ek(t)) = ek(t).
reduc isenc(enc(ek(t1),t2,t3)) = enc(ek(t1),t2,t3);
      isenc(garbageEnc(ek(t1),t2)) = garbageEnc(ek(t1),t2).
reduc fst(pair(x,y)) = x.
reduc snd(pair(x,y)) = y.
reduc ekof(enc(ek(t1),m,t2)) = ek(t1);
      ekof(garbageEnc(t1,t2)) = t1.
reduc equal(x,x) = x.
reduc verify(vk(t1),sig(sk(t1),t2,t3)) = t2.
reduc issig(sig(sk(t1),t2,t3)) = sig(sk(t1),t2,t3);
      issig(garbageSig(t1,t2)) = garbageSig(t1,t2).
reduc vkof(sig(sk(t1),t2,t3)) = vk(t1);
      vkof(garbageSig(t1,t2)) = t1.
reduc isvk(vk(t1)) = vk(t1).
reduc unstring0(string0(s)) = s.
reduc unstring1(string1(s)) = s.
reduc isek(ek(t)) = ek(t).
reduc isdk(dk(t)) = dk(t).
reduc ekofdk(dk(t)) = ek(t).
reduc issk(sk(t)) = sk(t).
reduc vkofsk(sk(t)) = vk(t).

query att:vs,pair(sl,sr).

let device =
    out(c, ek(k)) |
    ( ! lock; in(c, x); read s as y;
        if y = init then s := x; unlock ) |
    ( ! lock; in(c, x); read s as y;
        let z = dec(dk(k),x) in
        let zl = fst(z) in
        let zr = snd(z) in
        if y = left then out(c, zl); unlock
        else if y = right then out(c, zr); unlock ).
let user =
    new sl; new sr; new r;
    out(c, enc(ek(k), pair(sl, sr), r)).
process
    new k; new s; [s |-> init] | device | ! user

\end{lstlisting}
\newpage
\section*{Appendix D: Left-or-right Protocol in SAPIC Syntax}

\begin{lstlisting}[language=C]

theory LeftRightCase
begin

functions:
enc/3, ek/1, dk/1, sig/3, vk/1, sk/1, pair/2, string0/1,
string1/1, empty/0, garbageSig/2, garbage/1, garbageEnc/2
dec/2, isenc/1, isek/1, isdk/1, ekof/1, ekofdk/1, verify/2,
issig/1, isvk/1, issk/1, vkof/1, vkofsk/1, fst/1, snd/1,
unstring0/1, unstring1/1

equations:
dec(dk(t1), enc(ek(t1), m, t2)) = m,
isenc(enc(ek(t1), t2, t3)) = enc(ek(t1), t2, t3),
isenc(garbageEnc(t1, t2)) = garbageEnc(t1, t2),
isek(ek(t)) = ek(t),
isdk(dk(t)) = dk(t),
ekof(enc(ek(t1), m, t2)) = ek(t1),
ekof(garbageEnc(t1, t2)) = t1,
verify(vk(t1), sig(sk(t1), t2, t3)) = t2,
issig(sig(sk(t1), t2, t3)) = sig(sk(t1), t2, t3),
issig(garbageSig(t1, t2)) = garbageSig(t1, t2),
isvk(vk(t1)) = vk(t1),
issk(sk(t1)) = sk(t1),
vkof(garbageSig(t1, t2)) = t1,
fst(pair(x, y)) = x,
snd(pair(x, y)) = y,
unstring0(string0(s)) = s,
unstring1(string1(s)) = s

let Device=(
    out(ek(sk))
    ||
    (   in(req);
        lock s ;
        lookup s as ys in
            if ys='init' then
                insert s,req;
                unlock s
            else unlock s
    )
    ||
    (
        lock s;
        in(x);
        if isenc(x) = x then
        if ekof(x) = ek(sk) then
        if pair(fst(dec(dk(sk), x)), snd(dec(dk(sk), x)))
                    = dec(dk(sk), x) then
        lookup s as y in
            if y='left' then
                event Access(fst(dec(dk(sk), x)));
                            out(fst(dec(dk(sk), x))); unlock s
            else if y='right' then
                event Access(snd(dec(dk(sk), x)));
                            out(snd(dec(dk(sk), x))); unlock s
        else unlock s
        )
)

let User=new lm; new rm; new rnd; event Exclusive(lm,rm);
         out(enc(ek(sk), pair(lm, rm), rnd))

!( new sk; new s; insert s,'init'; ( Device || ! User ))


lemma types [typing]:
 All m #i. Access(m)@i ==>
    (Ex #j. KU(m)@j & j<i)
    |(Ex x #j. Exclusive(x,m)@j)
    |(Ex y #j. Exclusive(m,y)@j)

lemma secrecy:
 not(Ex x y #i #k1 #k2. Exclusive(x,y)@i & K(x)@k1 & K(y)@k2)

end

\end{lstlisting}

\end{document}